\documentclass[showpacs,preprintnumbers,superscriptaddress,nofootinbib]{revtex4}

\usepackage{color}
\usepackage{graphicx}
\usepackage{axodraw}

\def\ls#1{\ifmath{_{\lower1.5pt\hbox{$\scriptstyle #1$}}}}
\def\lss#1{\ifmath{^{\,\lower2.5pt\hbox{$\scriptstyle #1$}}}}

\def\sig{\sigma}

\def\lamT{\lambda_T}
\def\lamU{\lambda_U}

\def\cbma{c_{\beta-\alpha}}

\def\sbma{s_{\beta-\alpha}}

\def\stwob{s_{2\beta}}

\def\lamtil{\lam\ls{345}}

\def\cbma{c_{\beta-\alpha}}

\def\sbma{s_{\beta-\alpha}}

\def\stwob{s_{2\beta}}

\def\lamtil{\lam_{345}}

\def\tb{t_{\beta}}

\def\sb  {s_{\beta}}
\def\cb  {c_{\beta}}
\def\stwob  {s_{2\beta}}

\def\sa  {s_{\alpha}}
\def\ca  {c_{\alpha}}

\def\tanb{\tan\beta}
\def\cotb{\cot\beta}

\def\mz{m_Z}

\def\lam{\lambda}

\def\cala{{\cal A}}
\def\calm{{\cal M}}

\def\nn{\nonumber}

\def\wpm{W^{\pm}}
\def\hpm{H^{\pm}}

\def\gmp{G^{\mp}}

\def\calm{{\cal M}}
\def\wtil{\widetilde}
\def\what{\widehat}

\def\lsim{\mathrel{\raise.3ex\hbox{$<$\kern-.75em\lower1ex\hbox{$\sim$}}}}
\def\gsim{\mathrel{\raise.3ex\hbox{$>$\kern-.75em\lower1ex\hbox{$\sim$}}}}
\def\ifmath#1{\relax\ifmmode #1\else $#1$\fi}
\def\half{\ifmath{{\textstyle{1 \over 2}}}}

\def\quarter{\ifmath{{\textstyle{1 \over 4}}}}

\def\vev#1{\langle #1 \rangle}
\def\lam{\lambda}

\def\gl{\wt g}
\def\mgl{m_{\gl}}

\def\calo{{\cal O}}

\def\etal{{\it et al.}}

\def\stopl{\wt t_L}

\def\mstopl{m_{\stopl}}

\def\sbotl{\wt b_L}

\def\msbotl{m_{\sbotl}}

\def\msusy{m_{\rm SUSY}}

\def\calo{{\cal O}}

\def\etal{{\it et al.}}

\def\gl{\wt g}
\def\mgl{m_{\gl}}

\def\hl{h^0}
\def\hh{H^0}
\def\ha{A^0}
\def\hp{H^+}
\def\hm{H^-}
\def\gp{G^+}
\def\gm{G^-}
\def\go{G^0}
\def\hpm{H^{\pm}}
\def\hmp{H^{\mp}}
\def\mhl{m_{\hl}}
\def\mhh{m_{\hh}}
\def\mha{m_{\ha}}

\def\mhpm{m_{\hpm}}
\def\tanb{\tan\beta}
\def\cotb{\cot\beta}
\def\mt{m_t}
\def\mb{m_b}
\def\mz{m_Z}
\def\mw{m_W}

\def\wp{W^+}
\def\wm{W^-}
\def\wpm{W^{\pm}}

\def\wt{\widetilde}

\def\MPL #1 #2 #3 {{\sl Mod.~Phys.~Lett.}~{\bf#1} (#3) #2}
\def\NPB #1 #2 #3 {{\sl Nucl.~Phys.}~{\bf #1} (#3) #2}
\def\PLB #1 #2 #3 {{\sl Phys.~Lett.}~{\bf #1} (#3) #2}
\def\PR #1 #2 #3 {{\sl Phys.~Rep.}~{\bf#1} (#3) #2}
\def\PRD #1 #2 #3 {{\sl Phys.~Rev.}~{\bf #1} (#3) #2}
\def\PRL #1 #2 #3 {{\sl Phys.~Rev.~Lett.}~{\bf#1} (#3) #2}
\def\RMP #1 #2 #3 {{\sl Rev.~Mod.~Phys.}~{\bf#1} (#3) #2}
\def\ZPC #1 #2 #3 {{\sl Z.~Phys.}~{\bf #1} (#3) #2}
\def\IJMP #1 #2 #3 {{\sl Int.~J.~Mod.~Phys.}~{\bf#1} (#3) #2}
\def\NIM #1 #2 #3 {{\sl Nucl.~Inst.~and~Meth.}~{\bf#1} {#3} #2}

\def\lam{\lambda}
\def\br{B}

\def\gam{\gamma}

\def\etal{{\it et al.}}

\def\anti{\overline}
\def\epem{e^+e^-}

\def\rts{\sqrt s}

\def\eps{\epsilon}
\def\anti{\overline}

\def\gamha{\Gamma_{\ha}^{\rm tot}}

\def\fbi{~{\rm fb}^{-1}}

\def\gev{~{\rm GeV}}
\def\tev{~{\rm TeV}}
\def\mt{m_t}
\def\mb{m_b}

\newcommand{\nc}{\newcommand}
\nc{\beq}{\begin{equation}}   \nc{\eeq}{\end{equation}}
\nc{\bea}{\begin{eqnarray}}   \nc{\eea}{\end{eqnarray}}
\nc{\baa}{\begin{array}}      \nc{\eaa}{\end{array}}
\nc{\bit}{\begin{itemize}}    \nc{\eit}{\end{itemize}}
\nc{\ben}{\begin{enumerate}}  \nc{\een}{\end{enumerate}}
\nc{\bce}{\begin{center}}     \nc{\ece}{\end{center}}
\def\beqa{\begin{eqnarray}}
\def\eeqa{\end{eqnarray}}
\def\bed{\begin{description}}
\def\eed{\end{description}}

\def\gl{\wt g}
\def\mgl{m_{\gl}}

\def\calo{{\cal O}}

\def\etal{{\it et al.}}

\def\rta{\rightarrow}
\def\tanb{\tan\beta}

\def\lsim{\mathrel{\raise.3ex\hbox{$<$\kern-.75em\lower1ex\hbox{$\sim$}}}}
\def\gsim{\mathrel{\raise.3ex\hbox{$>$\kern-.75em\lower1ex\hbox{$\sim$}}}}

\def\process{e^+e^-\to\nu\anti\nu \ha}
\def\sp{\what s}
\def\ep{e^+}
\def\nn{\nu}
\def\nc{\overline\nu}

\begin{document}
\bibliographystyle{revtex}

\preprint{CALT-68--2429}
\preprint{MADPH--03--1323}
\preprint{UCD-03-1}
\preprint{hep-ph/0302266}

\title{$e^+e^- \to \nu \anti \nu \ha$ in the two-Higgs-doublet model}

\author{Tom Farris}
\email{farris@physics.ucdavis.edu}
\affiliation{Davis Institute for High Energy Physics,
University of California, Davis, CA 95616, USA}

\author{John F. Gunion}
\email{jfgucd@physics.ucdavis.edu}
\affiliation{Davis Institute for High Energy Physics,
University of California, Davis, CA 95616, USA}

\author{Heather E. Logan}
\email{logan@pheno.physics.wisc.edu}
\affiliation{Department of Physics, University of Wisconsin, 1150 University
Avenue, Madison, WI 53706, USA}

\author{Shufang Su}
\email{shufang@theory.caltech.edu}
\affiliation{California Institute of Technology, Pasadena, 
California 91125, USA}

\date{\today}

\begin{abstract}
We compute the cross section for $e^+e^-\rightarrow{\nu}\anti{\nu}\ha$ in
the general CP-conserving type-II two-Higgs-doublet model.
We sum the contributions from 
the ``$t$-channel'' $\epem\to \nu\anti\nu WW\to \nu\anti\nu\ha$ graphs
and ``$s$-channel'' $\epem\to Z \ha\to \nu\anti\nu \ha$ graphs, 
including their interference. 
Higgs-triangle graphs and all box diagrams are included. For
many parameter choices, especially those in the decoupling region
of parameter space (light $\hl$ and $\mha,\mhh,\mhpm>2\mz$) the Higgs-triangle
and box diagrams are found
to be of minor importance, the main contributing
loops being the top and bottom quark triangle diagrams. The predicted
cross section is rather small for $\tanb>2$ and/or $\mha>2m_t$.
However, we also show that
if parameters are chosen corresponding to large Higgs self-couplings
then the Higgs-triangle graphs can greatly enhance the cross section.
We also demonstrate that the SUSY-loop corrections to the $b\anti b\ha$
coupling could be such as to greatly enhance this coupling, resulting
in an enhanced $\nu\anti\nu\ha$ cross section.
Complete cross section expressions are given in the Appendices. 
\end{abstract}

\pacs{12.60.Jv, 12.60.Fr, 14.80.Cp, 14.80.Ly}

\maketitle


\section{\label{sec:intro}Introduction}
The Higgs mechanism provides an elegant way to explain 
electroweak symmetry breaking (EWSB) and the origin of the 
masses of all the observed Standard Model (SM) particles. 
In many approaches, the symmetry breaking arises
from a sector involving scalar fields and leaves behind
one or more physical Higgs bosons.
Detecting and studying all such Higgs bosons
is one of the major objectives of current and 
future particle physics experiments. 
The minimal SM contains one 
${\rm SU(2)}_{\rm L}$-doublet Higgs field, 
leading to a single physical CP-even 
Higgs boson after EWSB.  However, the 
electroweak scale is not stable with respect to radiative corrections
in the minimal SM. Numerous extensions of the SM have been
proposed to cure this naturalness/hierarchy problem, many of which
predict a low-energy effective theory with a Higgs sector that
contains two (or more) Higgs doublet fields. Our focus here will
be on a general two-Higgs-doublet model (2HDM)
(for a review and references see \cite{HHG}).
In particular, the most promising
extension of the SM that resolves the naturalness
and hierarchy problems is low-energy supersymmetry (SUSY), which
must contain at least two Higgs doublets.  For precise gauge-coupling
unification, exactly two doublets are preferred, as incorporated
in the minimal supersymmetric Standard Model (MSSM). 
The two-doublet Higgs sector of the MSSM 
is predicted to be CP-conserving at tree level
and to have type-II fermionic couplings in which one
doublet ($\Phi_1$) gives mass to down-type quarks and charged leptons while
the other doublet ($\Phi_2$) gives mass to up-type quarks.
In this case there are five physical Higgs bosons: 
the CP-even $\hl$ and $\hh$, the CP-odd  $\ha$, and the charged 
pair $H^{\pm}$. The most important additional parameters
of the CP-conserving type-II 
2HDM are: (i) $\tanb$ ($\tanb\equiv v_2/v_1=\vev{\Phi_2^0}/\vev{\Phi_1^0}$,
the ratio of the vacuum expectation values 
of the neutral members of the two Higgs doublets);
and (ii) the mixing angle $\alpha$
that diagonalizes the neutral CP-even Higgs sector.

Higgs searches at the CERN LEP II collider 
have excluded a SM Higgs boson with mass below
114.4 GeV \cite{lep} at 95\% confidence level. 
In the context of specific choices for the 
soft-SUSY-breaking parameters at the TeV scale,
LEP II can be used to exclude 
a range of MSSM Higgs masses and $\tanb$. For example, 
assuming the maximal-top-squark-mixing
scenario with $\msusy=1\tev$, LEP II excludes $\mhl<91.0$ GeV,
$\mha<91.9$ GeV  and $0.5<\tanb<2.4$ at 95\% Confidence 
Level (CL)~\cite{lepa,lep2}.
Searches for top quark decay $t\to H^+b$ at the Fermilab Tevatron
exclude $\tanb>50$ when $m_t>\mhpm$, and searches for the final
state $b\anti b h^0 \to b\anti b b\anti b$ exclude very high values of $\tanb$
as a function of the $h^0$ mass \cite{tevsum}. 
Precision electroweak measurements provide only weak
constraints on $\tan\beta$ \cite{yamada}.
Finally, limits can be placed on $\tanb$ as a function of $\mhpm$
based on $\pi^+,K^+,B^+\to\mu^+\nu$ decays 
and on $K$-$\anti K$ and $B$-$\anti B$
mixing \cite{Marin:2002}. In the context of a type-II model, 
as defined earlier, 
these roughly require $1<\tanb<200$ for $\mhpm\sim 300\gev$.
However, much of the MSSM parameter space remains to be explored at future
collider experiments. 
Indeed,  for the most general MSSM boundary
conditions, there is no lower bound on $\tanb$ from LEP II data.
For the most general 2HDM, only the presence of two 
simultaneously light Higgs bosons can be 
excluded~\cite{Grzadkowski:1999ye,opal2hdm}.

At Run II of the Tevatron,
discovery of the light CP-even Higgs boson
$\hl$ of the MSSM at the $5\sigma$ 
level is possible for $\mhl\lsim 120$ GeV with 15 ${\rm fb}^{-1}$ 
of integrated luminosity \cite{tev}.
At the CERN Large Hadron Collider (LHC), 
discovery of $\hl$ is virtually guaranteed over all of the 
MSSM parameter space \cite{lhcATLAS,lhcCMS,perini}, and measurements of
ratios of the
the $\hl$ partial decay widths in the more prominent decay channels
will be possible with precisions on the order of $15\%$  \cite{Zeppenfeld}.  
A linear $e^+e^-$ collider could 
make precision measurements of the $\hl$ couplings with accuracies
of a few percent \cite{hLCmeas,TeslaTDR,Orangebook}. 
In fact, at a linear collider with $\rts\geq 350\gev$, at least
one of the CP-even Higgs bosons of a general 2HDM (or
more complicated Higgs sector) is {\it guaranteed} to be 
detected~\cite{Espinosa:1998xj} in the $Zh$ production mode. 
In contrast, even in the simple 2HDM
there is no guarantee that the CP-odd $\ha$ can be detected,
and the situation only worsens
for more complicated Higgs sectors. Thus, it behooves us to
explore every option for $\ha$ production.

For a CP-conserving Higgs sector, production
of a single $\ha$ via loop-induced processes could prove critical
to a full exploration of the Higgs sector. This is because
the most useful tree-level mechanisms for single Higgs production
rely on a substantial Higgs coupling to $ZZ$ or $WW$ pairs.  Such
couplings are absent at tree-level for the purely CP-odd $\ha$.
If the Higgs sector is CP-violating, then the neutral Higgs bosons
will mix with one another and, in general, all will have substantial
tree-level $ZZ$ and $WW$ couplings.
As a result, in the case of a CP-violating Higgs sector,
loop-induced $\ha$ production mechanisms would probably
not be very important.
Thus, loop-induced $\ha$ production is mainly of interest for a 
CP-conserving Higgs sector.
We note that there are substantial phenomenological reasons
for believing that the Higgs sector will prove to be CP-conserving.
In particular, CP-conservation is the most straightforward approach to avoiding
conflict with the constraints coming from the anomalous
magnetic moment of the muon $(g_\mu-2)$~\cite{Bennett:2002jb} 
and the non-observation of electric dipole moments (EDMs). 
Still, we must note that even in the MSSM context
substantial CP violation could be introduced 
at the loop level if the soft-SUSY-breaking parameters have phases, and
that a CP-violating Higgs sector can be consistent with
the EDM and $(g_\mu-2)$ constraints if there are carefully orchestrated
cancellations between CP-violating contributions to these 
observables~\cite{cpcancellations}. 

Let us review in detail the difficulties associated
with producing and detecting a purely CP-odd $\ha$.
At a hadron collider, the absence of tree-level
$ZZ\ha$ and $WW\ha$ couplings implies that: 
(i) the $W^*\to W\ha$ production
mode is suppressed, as particularly relevant at the Tevatron; (ii) the
$gg\to\ha\to WW$ and $gg\to \ha\to ZZ$ production/decay modes 
(the ``gold-plated'' processes for discovery of a heavy SM-like Higgs boson)
have very low rates because the branching fractions
$\br(\ha\to ZZ)$ and $\br(\ha\to WW)$
are small; and (iii)  at the LHC, the $gg\to\ha\to \gam\gam$
rate is numerically small. With regard to the latter, we note that
had $\Gamma(\ha\to \gam\gam)$ been of SM-like size, the absence
of tree-level $\ha\to WW,ZZ$ decays would have implied substantial
$\br(\ha\to\gam\gam)$ and a useful
$gg\to\ha\to\gam\gam$ rate even for large $\mha$.  However,
the absence of the $W$-loop (the largest contribution in the case of
a SM-like Higgs) results in an 
even greater suppression of $\Gamma(\ha\to\gam\gam)$ than of $\gamha$.
At an $\epem$ collider, the
$\epem\to Z^*\to Z\ha$ and $\epem\to \nu\anti\nu W^*W^*\to\nu\anti\nu \ha$
processes are only present at the one-loop level.
 
In general, the $\ha$ {\it can} be pair produced at tree-level.
However, 
the rates for Higgs pair production are generally too small for observation
at the Tevatron and LHC since they are electroweak in strength
and must compete against enormous QCD backgrounds.
Pair production is, however, potentially
useful at a $\epem$ machine.  Such processes include
$\epem\to Z^*\to Z\ha\ha$ \cite{zzaa,Farris:2002ny},
$\epem\to \nu\anti \nu W^*W^*\to 
\nu\anti \nu \ha\ha$~\cite{wwaa,Farris:2002ny}
and $\epem\to Z^*\to \hh\ha$ or $\hl\ha$
(see \cite{HHG}). However, all of
these processes can be simultaneously suppressed by kinematics 
and/or small couplings. In particular, this occurs
in the decoupling limit of a 2HDM that typically
arises when $\mha>2\mz$~\cite{Gunion:2002zf}. In the decoupling limit
$\mhh\sim\mha\gg\mhl$, implying that $\rts>2\mha$ is required for $\hh\ha$,
$Z\ha\ha$ and $\nu\anti\nu \ha\ha$ production
(all of which would otherwise have large cross sections
since the $ZZ\ha\ha$ and $WW\ha\ha$ couplings are fixed by the standard 
quadratic gauge couplings of the $\ha$ and the $Z\hh\ha$
coupling is maximal in this limit), while $\hl\ha$ production
is strongly suppressed in the decoupling limit by 
a factor of order $\mz^4/\mha^4$ in the square of the $\hl\ha Z$ coupling.
This decoupling limit is automatic in the context of the MSSM and
is quite natural in the case of a more general 2HDM. 
In the general 2HDM there are other scenarios not related
to this standard decoupling limit in which detection of the $\ha$
on its own would also be critical. In particular, it is possible to choose
Higgs sector parameters in such a way that the $\ha$ is the
only light Higgs boson while maintaining consistency with
precision electroweak measurements~\cite{Chankowski:2000an} (see also
\cite{Gunion:2002zf}).  Further, it is possible that a light $\ha$
could explain part of the observed discrepancy between the
SM prediction for $(g_\mu-2)$ and the experimentally measured 
value~\cite{Cheung:2001hz}.
For all the above reasons, it is important to assess more carefully
the various possible mechanisms for single $\ha$ production.

Consider first the tree-level processes for single $\ha$ production.
At both the LHC and at an $\epem$ collider, the only
relevant tree-level processes for single $\ha$ production
are $t\anti t \ha$ and $b\anti b \ha$ production. 
At the LHC, it has long been established
that these do not yield an
observable signal in a wedge-shaped region of the $(\mha,\tanb)$
parameter space. This wedge spans a range of moderate $\tanb$
values for $\mha>200\gev$ and becomes increasingly broad 
as $\mha$ increases (see, for example, the ATLAS and CMS TDRs
\cite{lhcATLAS,lhcCMS}).  At an $\epem$ collider, this wedge is even larger,
beginning at a relatively low value of $\mha$ 
(the precise value depends on the $\rts$ of the 
machine)~\cite{Djouadi:gp,Grzadkowski:1999ye,Grzadkowski:2000wj}
(see also \cite{bbtta3,bbtta4}).
In particular, for $\mha+2m_t>\rts$
(and if Higgs pair production 
processes are suppressed or kinematically forbidden)
there is no known means for detecting the $\ha$ 
using tree-level production mechanisms when $\tanb$ is not large
enough for the $b\anti b \ha$ process to have an observable rate.
In this situation, we must turn to loop-induced $\ha$ 
production mechanisms.

One possibility is to build a photon collider at the $\epem$ 
collider and look for $\gam\gam\to \ha$ via $t$, $b$ and
charged Higgs loops~\cite{Gunion:1992ce,Asner:2001ia,Muhlleitner:2001kw}. 
In particular, the recent realistic study of Ref.~\cite{Asner:2001ia}
found that three years of running at a 630 GeV $e^+e^-$ LC in 
the photon collider mode could provide a 4$\sigma$ signal 
for a significant fraction of the $(\mha,\tanb)$ LHC wedge region
with $\mha\leq500\gev$.  
If $\mha$ could be roughly guessed, e.g., based on 
the precise measurements of deviations of the $h^0$ couplings
from their SM 
values~\cite{Battaglia:2000jb,Carena:2001bg,TeslaTDR,Kiyoura:2001kj,ACFA}
obtainable at a linear $e^+e^-$ collider,
then the energy of the photon 
collider could be chosen to optimize the $\ha$ production cross section
resulting in a faster discovery.
One-loop production possibilities in $\epem$ collisions
include $\epem\to \gam \ha$, $\epem\to Z^*\to Z\ha$ and 
$\epem\to \nu\anti\nu\ha$
via $WW$ fusion. (The latter two interfere since
$Z\to\nu\anti\nu$.) The $\epem\to\gam\ha$ process has been computed
in Refs.~\cite{Djouadi:1996ws,Akeroyd:1999gu}.  
Results for the $Z\ha$ final state (computed assuming
the absence of supersymmetric particle loops) appear in 
Refs.~\cite{Akeroyd:1999gu,Farris:2002ny}. The possible enhancement
of the $Z\ha$ rate when SUSY particles are present is discussed
in \cite{eeZAMSSM}. 

In this paper, we give a complete calculation of the 
$W$ boson fusion process $\epem\to \nu\anti\nu W^*W^*\to \nu\anti\nu\ha$
in the general CP-conserving 2HDM,
which first occurs at one-loop since there is no tree level 
$W^+W^-\ha$ coupling.
We include the process $\epem\to Z^*\to Z\ha\to\nu\anti\nu\ha$
(first computed in Ref.~\cite{Akeroyd:1999gu}),
which leads to the same final state and thus interferes with the
$W$ boson fusion process.
The process $e^+e^- \to \nu \bar \nu \ha$ has also been computed
recently in Ref.~\cite{Arhrib:2002ti}.
After a review of the structure of the general 2HDM in Sec.~\ref{sec:2HDM},
we present the relevant Feynman diagrams and 
formulae involved in our calculations in Sec.~\ref{sec:formalism}.
We present numerical results in Sec.~\ref{sec:numerics}.

In Sec.~\ref{subsec:MSSM2HDM}, we employ the tree-level 
MSSM two-doublet sector (see \cite{HHG}) as a benchmark
for our study. The Higgs sector of the MSSM is a type-II 2HDM in which 
the quartic couplings of the two Higgs doublet fields
are fixed at tree level by the gauge couplings. In this case,
all Higgs sector parameters are determined by just the two
parameters $\mha$ and $\tanb$, and for $\mha>\mz$ the 2HDM quickly approaches
the decoupling limit, described earlier, in which all Higgs self-couplings
are small. 
We compare the full 2HDM results including the loop contributions from 
quarks, Higgs and gauge bosons with those obtained by including
only the top and bottom quark loops, as computed in \cite{Farris:2002ny}.
We show that the $\epem\to \nu\anti\nu\ha$ process 
could provide a viable $\ha$ signal for $\tanb<1$, 
thus covering part of the region where $\ha$ discovery using
tree-level processes is not possible.
Further, for such $\tanb$ values the $\nu\anti\nu \ha$ rate would 
provide a very sensitive measurement of $\tanb$. 
In contrast to the above, the $\nu\anti\nu\ha$
cross section is typically quite small for $\tanb\geq 2$ for
parameter choices based on the tree-level MSSM Higgs sector potential.
Of course, in the full MSSM, radiative corrections to the tree-level
masses should be incorporated as should the contributions with 
superparticles running in the loop.   
However, the sparticle loops are in general suppressed by the 
heavy sparticle masses.  A full study of the MSSM,
including all the superparticle loop contributions and radiative
corrections to the Higgs potential, will appear in \cite{nunuafull}.

Sec.~\ref{subsec:2HDM} focuses on the general 2HDM 
with Higgs potential parameters outside the decoupling regime. 
We find that if the Higgs self-couplings are large
(as possible when 2HDM parameters
are chosen to lie in a non-decoupling regime)
then the $\nu\anti\nu\ha$ cross section can be greatly enhanced by
Higgs triangle diagrams even for such large $\hl,\hh,\hpm$ masses
that none of these latter Higgs bosons could be directly observed.
Indeed, for lower $\mha$ values ($\mha<2\mt$), 
the rate is sufficiently enhanced when $\tanb>10$
that the $\nu\anti\nu \ha$ events could have a detectable
rate and provide direct evidence
for the large Higgs self-couplings.  This probe
of the Higgs self-couplings would be especially powerful
if some of the other Higgs bosons have themselves been directly observed.

In Sec.~\ref{subsec:MSSMspecial}, we illustrate one unusual
possibility in the full MSSM context, namely 
that the $b\anti b\ha$ coupling could be greatly enhanced by
non-decoupling SUSY particle loops, resulting
in a huge enhancement for the $\epem\to\nu\anti\nu\ha$
cross section. 

Finally, Sec.~\ref{sec:conclusions} is reserved for our conclusions.  In the 
Appendices, we collect the complete matrix elements for the various Feynman 
diagram contributions.

\section{\label{sec:2HDM}The CP-conserving 2HDM}

We adopt the conventions of \cite{Gunion:2002zf} for the 2HDM. 
Let $\Phi_1$ and
$\Phi_2$ denote two complex SU(2)$_{L}$-doublet scalar fields
with hypercharge $Y=1$.
The most general gauge invariant scalar potential is given by
\bea
\mathcal{V}&=&m_{11}^2\Phi_1^\dagger\Phi_1+m_{22}^2\Phi_2^\dagger\Phi_2
-[m_{12}^2\Phi_1^\dagger\Phi_2+{\rm h.c.}]
+\half\lambda_1(\Phi_1^\dagger\Phi_1)^2
+\half\lambda_2(\Phi_2^\dagger\Phi_2)^2
+\lambda_3(\Phi_1^\dagger\Phi_1)(\Phi_2^\dagger\Phi_2)
+\lambda_4(\Phi_1^\dagger\Phi_2)(\Phi_2^\dagger\Phi_1)
\nonumber\\[6pt]
&&\quad 
+\left\{\half\lambda_5(\Phi_1^\dagger\Phi_2)^2
+\big[\lambda_6(\Phi_1^\dagger\Phi_1)
+\lambda_7(\Phi_2^\dagger\Phi_2)\big]
\Phi_1^\dagger\Phi_2+{\rm h.c.}\right\}\,. \label{pot}
\eea
The terms proportional to $\lambda_6$ and $\lambda_7$ lead to
flavor-changing neutral current interactions (FCNCs) and 
will be set to zero.  This can be achieved by
imposing a discrete symmetry $\Phi_1\rta -\Phi_1$ on ${\cal V}$.  
However, we allow for a soft (dimension-two) breaking of this symmetry 
through $m_{12}^2\neq 0$.\footnote{%
This discrete symmetry is also employed to restrict the Higgs-fermion couplings
so that no tree-level Higgs-mediated FCNCs are present.} If
$\lam_6=\lam_7=0$ but $m_{12}^2\neq 0$, the soft breaking
of the discrete symmetry generates {\it finite} Higgs-mediated
FCNCs at one loop.  The tree-level supersymmetric form of ${\cal V}$
is obtained from Eq.~(\ref{pot}) by the substitutions:
\bea
\lambda_1 =\lambda_2 = \quarter (g^2+g'^2)\,, \quad 
\lambda_3 =\quarter (g^2-g'^2)\,,   \quad
\lambda_4 =-\half g^2\,, \quad 
\lambda_5 =\lambda_6=\lambda_7=0\,,\label{bndfr}
\eea
where $g$ and $g^{\prime}$ are the ${\rm SU(2)}_L$ and ${\rm U(1)}_Y$
gauge couplings, respectively. In general, $m_{12}^2$ and $\lambda_5$
(and, if present, $\lambda_6$ and $\lambda_7$) can be complex.  However, we
explicitly exclude such CP-violating phases
by choosing all coefficients in Eq.~(\ref{pot}) to be
real and such that spontaneous CP violation is absent.  
For details, see Ref.~\cite{Gunion:2002zf}.
The scalar fields will
develop non-zero vacuum expectation values if the mass matrix
has at least one negative eigenvalue. Imposing CP invariance
and U(1)$_{\rm EM}$ gauge symmetry, the minimum of the potential
corresponds to the following vacuum expectation values:
\beq
\langle \Phi_1 \rangle={1\over\sqrt{2}} \left(
\begin{array}{c} 0\\ v_1\end{array}\right), \qquad \langle
\Phi_2\rangle=
{1\over\sqrt{2}}\left(\begin{array}{c}0\\ v_2
\end{array}\right)\,,\label{potmin}
\eeq
where $v_i$ are real in the absence of explicit and spontaneous CP violation.
The minimization conditions on the potential can then be used to 
determine $m_{11}^2$
and $m_{22}^2$ in terms of the other parameters (with $\lam_6=\lam_7=0$):
\bea
m_{11}^2 = m_{12}^2\tb -\half
v^2\left[\lam_1\cb^2+\lamtil\sb^2\right]
\,,\quad
m_{22}^2 = m_{12}^2\tb^{-1}-\half v^2
\left[\lam_2\sb^2+\lamtil\cb^2\right]\,,
\label{minconditions}
\eea
where we have defined:
\beq
\lamtil\equiv\lam_3+\lam_4+\lam_5\,,\qquad\qquad\tb\equiv\tanb\equiv{v_2\over
  v_1}\,,
\label{tanbdef}
\eeq
and $v^2\equiv v_1^2+v_2^2=4\mw^2/ g^2=(246~{\rm GeV})^2\,.$
It is always possible to choose the phases 
of the Higgs doublet fields such that
both $v_1$ and $v_2$ are positive, implying that we
can take $0\leq\beta\leq\pi/2$.
With $\lam_6=\lam_7=0$, all but one of the eight free parameters in 
Eq.~(\ref{pot}) can be fixed after EWSB in terms of $v$,
$\tan\beta$, the four physical Higgs masses, 
and the mixing angle $\alpha$ 
required to diagonalize the neutral CP-even Higgs sector.
In our numerical analysis below, we use the parameter set
$\mha$, $\mhh$, $\mhl$, $\mhpm$, $\alpha$, $\tanb$ and $\lam_5$.
($v$ is of course fixed by the measured values of $m_W$ and $g$.)
The relations (D.20)--(D.23) of~\cite{Gunion:2002zf} 
with $\lam_6=\lam_7=0$ then give the other $\lam_i$ as:
\bea
\lam_1&=& {\mhh^2\ca^2+\mhl^2\sa^2-\mha^2\sb^2\over
  v^2\cb^2}-\lam_5\tb^2\,;\label{inverseA}\\
\lam_2&=& {\mhh^2\sa^2+\mhl^2\ca^2-\mha^2\cb^2\over
  v^2\sb^2}-\lam_5\tb^{-2}\,;\label{inverseB}\\
\lam_3&=& {(\mhh^2-\mhl^2)\sa\ca+(2\mhpm^2-\mha^2)\sb\cb\over
v^2\sb\cb}-\lam_5 \,;\label{inverseC}\\
\lam_4&=&{2(\mha^2-\mhpm^2)\over v^2}+\lam_5\,.\label{inverseD}
\eea
The mass parameters of the Higgs potential are given by
(D.17), (D.24) and (D.25) of~\cite{Gunion:2002zf}:
\bea
m_{12}^2&=&\sb\cb(\lam_5 v^2+\mha^2) \label{inverse5}\,;\\
m_{11}^2&=&
-{1\over 2\cb}\left(\mhh^2\ca\cbma-\mhl^2\sa\sbma\right)+\sb^2(\lam_5v^2+
\mha^2)\,;
\label{minconds1}
\\
m_{22}^2&=&
-{1\over 2\sb}\left(\mhl^2\ca\sbma+\mhh^2\sa\cbma\right)+\cb^2(\lam_5v^2+\mha^2)\,.
\label{minconds2}
\eea
In the supersymmetric limit,  
the $\lam_i$ are determined in terms of gauge couplings
as given in Eq.~(\ref{bndfr}) and the Higgs sector is then entirely
specified at tree-level by the two free parameters $\tanb$ and $\mha$. 

There are two types of 2HDM, depending on which Higgs field is responsible
for the masses of quarks and leptons.  In the type-I 2HDM, $\Phi_2$ 
gives masses to both quarks and leptons.
In the type-II 2HDM, $\Phi_2$ couples to up-type quarks while $\Phi_1$
couples to both down-type quarks and charged leptons. 
Consequently, the Yukawa couplings of the quarks and leptons
to the Higgs bosons are different in these two cases. For the $\ha$
we find $\mathcal{L} = i y_f \anti f \gamma_5 f \ha$ where:
\begin{eqnarray}
&&y_t=\frac{m_t}{v}\cot\beta\ \ \ 
y_b=\frac{m_b}{v}\cot\beta\ \ \ 
y_{\tau}=\frac{m_{\tau}}{v}\cot\beta\ \ \ {\rm Type\ I}.
\label{typei}
\\
&&y_t=\frac{m_t}{v}\cot\beta\ \ \ 
y_b=\frac{m_b}{v}\tan\beta\ \ \ 
y_{\tau}=\frac{m_{\tau}}{v}\tan\beta\ \ \ {\rm Type\ II}.
\label{typeii}
\end{eqnarray}
The MSSM is required to have type-II couplings and our analysis
will also assume type-II couplings for the general 2HDM case.

\section{\label{sec:formalism}Formalism}

\begin{figure}
\resizebox{\textwidth}{!}{\includegraphics*[0,261][610,717]{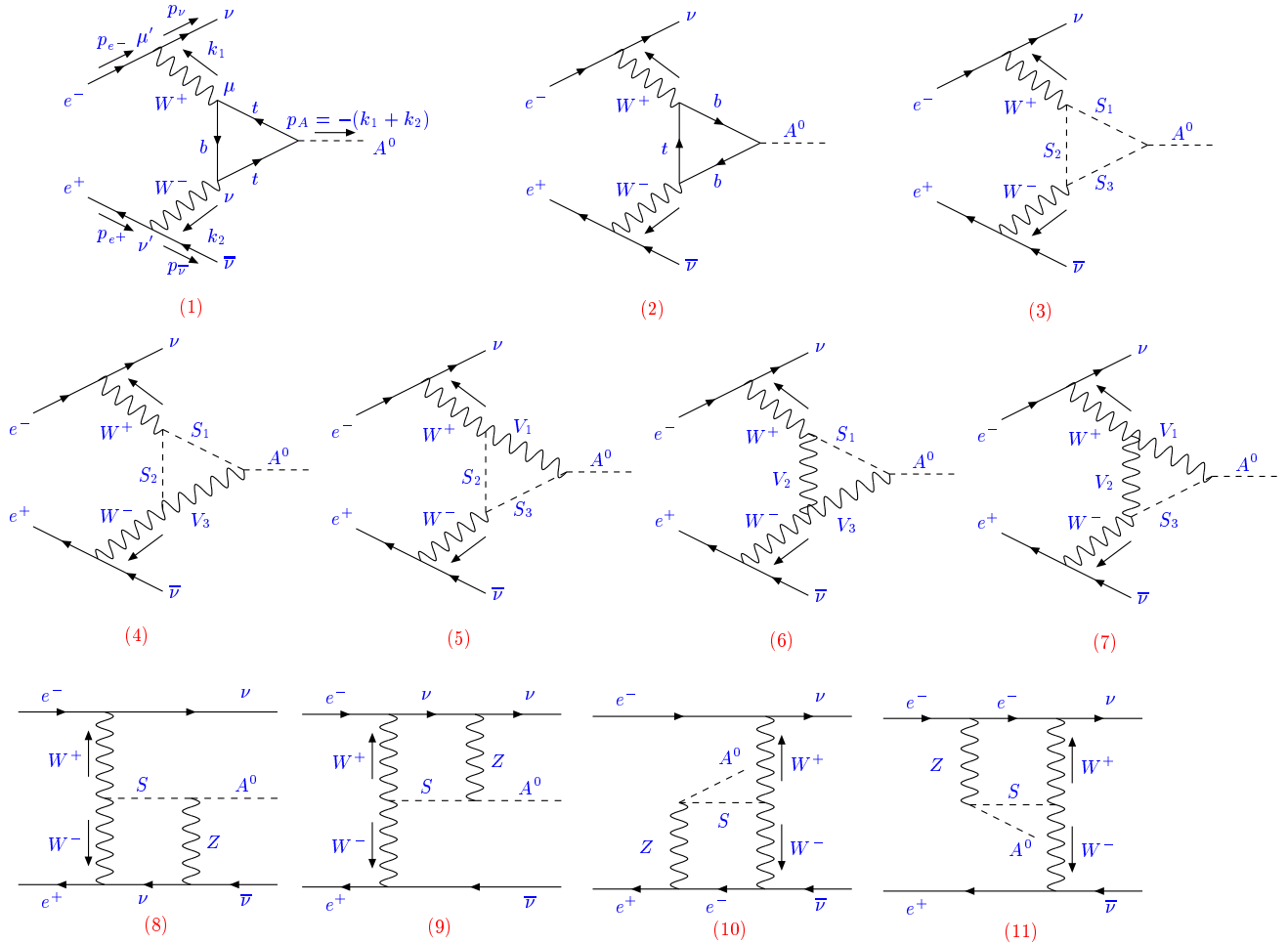}}
\caption{Feynman diagrams for 2HDM contributions to $e^+e^- \to
\nu\anti{\nu}\ha$ via $t$-channel processes. 
Here $S_i$ denotes Higgs and Goldstone bosons, 
$V_i$ denotes gauge bosons.}
\label{fig:WWA}
\end{figure}

In our analysis, we adopt the renormalization scheme of Ref.~\cite{cpr}.
The diagrams contributing to $e^+e^- \to \nu\anti{\nu}\ha$ in the 2HDM via 
$t$-channel processes are 
shown in Fig.~\ref{fig:WWA}, where we have neglected all the diagrams
that are proportional to the small electron Yukawa couplings. 
Diagrams (1)-(7) are the (finite) triangle loop corrections to 
the effective $\wp\wm\ha$ coupling, and diagrams 
(8)-(11) are the box diagrams. 
For on-shell $W$ bosons, the sum of diagrams (3)-(7) with Higgs bosons 
and vector bosons
running in the loop must exactly cancel \cite{gunhabkao}. 
But, in the present context,
the $W$ bosons are virtual and the sum of these diagrams is non-zero.
There is no $WW\ha$ counterterm contribution since the $WW\ha$ 
vertex is finite.  In addition, there are no tadpole contributions 
since we have set the renormalized tadpoles to be zero.
We shall refer to this collection of diagrams as the $t$-channel
diagrams. 

\begin{figure}
\includegraphics*[0,425][600,700]{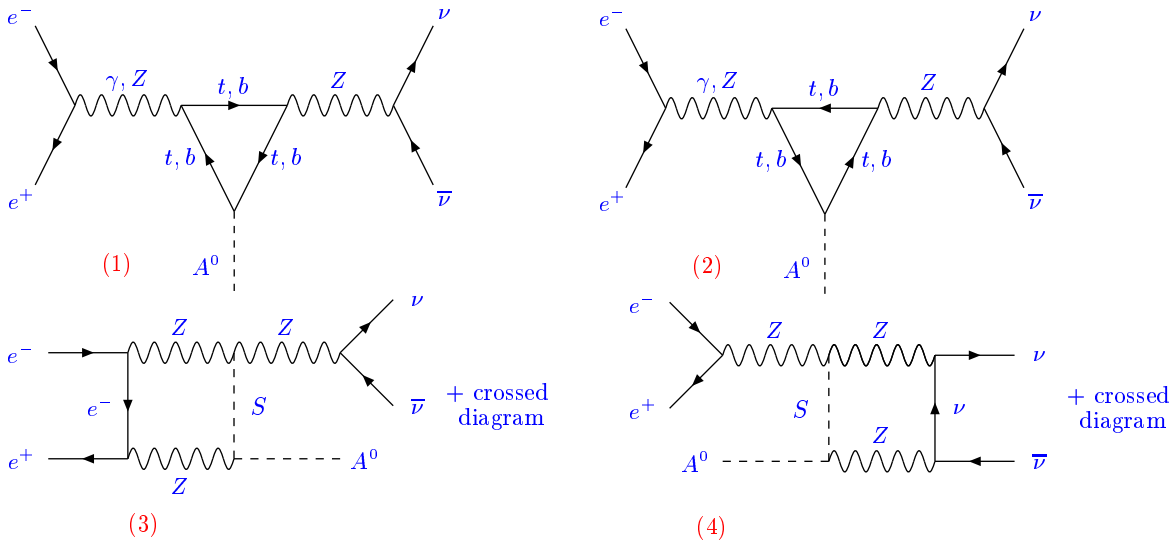}
\caption{Feynman diagrams for 2HDM contributions to $e^+e^- \to
\nu\anti{\nu}\ha$ via $s$-channel processes. }
\label{fig:ZZA}
\end{figure}

There are additional diagrams related to $s$-channel $\gam,Z$ exchange; 
these are shown in Fig.~\ref{fig:ZZA}.  
Diagrams (1) and (2) are the triangle loop contributions to the 
$\gamma Z\ha$ and $ZZA$ couplings (which are zero at tree-level), 
followed by $Z\to\nu\anti\nu$.  
Diagrams (3) and (4)  are related box diagrams; photon exchanges
do not appear in these diagrams because of the absence of $\gam\nu\anti\nu$
and $\gam Z S$ vertices.
If the final exchanged $Z$ connected to $\nu\anti \nu$ 
is on-shell in diagrams (1)-(3), then the calculation is equivalent 
to $e^+e^-\to Z \ha$. This process has been calculated for the 2HDM and 
the full MSSM in \cite{Akeroyd:1999gu, eeZAMSSM} and the 
production cross section found to be small. 
The $Z$ resonant contribution to the $\nu\anti\nu\ha$ final state
can be separated experimentally by detecting the $\ha$
in a visible final state (e.g. $b\anti b$), reconstructing
the mass of the $\nu\anti\nu$ recoiling opposite the $\ha$
and removing events in which the reconstructed mass is near $\mz$.
However, far off-shell intermediate $Z$ bosons can potentially give
$s$-channel contributions that interfere with the $t$-channel
contributions to the $\nu\anti\nu\ha$ final state, 
and must be included in our calculation.

One possible concern is that including the $Z$ decay width 
in the $Z$ propagators in the diagrams of Fig.~\ref{fig:ZZA} might
spoil the gauge invariance of the calculation.
We checked explicitly and found that this is not the case.
In particular, the sum of diagrams (1) and (2) 
in Fig.~\ref{fig:ZZA} is gauge invariant on its own. Also,
there is a cancellation between the 
gauge-dependent part in diagram (3) 
and the corresponding crossed diagram. 
A similar cancellation occurs between the gauge-dependent part of
diagram (4) and its crossed counterpart.
All these cancellations occur as a result of numerator algebra
and are independent of the $Z$ width appearing in the $Z$
propagators. 

Another check of the gauge independence of our fixed width scheme
for the $Z$ boson is to compare the numerical results to those of 
the ``factorization scheme'', which is guaranteed to be gauge independent.
Following, e.g., the discussion in Ref.~\cite{nnHrcs},
the one-loop matrix element in the factorization scheme is given 
by\footnote{For processes that are nonzero at tree level, care must be
taken to avoid double-counting the $Z$ width.  This is not a concern here
since the tree level matrix element is zero.}
\begin{equation}
        \mathcal{M} = \frac{\hat s - m_Z^2}{\hat s - m_Z^2 + i m_Z \Gamma_Z}
                \mathcal{M}_{\Gamma_Z = 0}.
\end{equation}
The factorization scheme sets the nonresonant diagram (4) of 
Fig.~\ref{fig:ZZA} to zero when $\hat s = M_Z^2$, which leads to 
an effect of $\mathcal{O}(\alpha \Gamma_Z / M_Z) \sim \mathcal{O}(\alpha^2)$;
since this is a higher-order effect, it should be small.
Our results in the fixed width scheme agree numerically with those of the 
factorization scheme to within the precision of our phase space integration.

\begin{figure}
\includegraphics*[50,320][570,650]{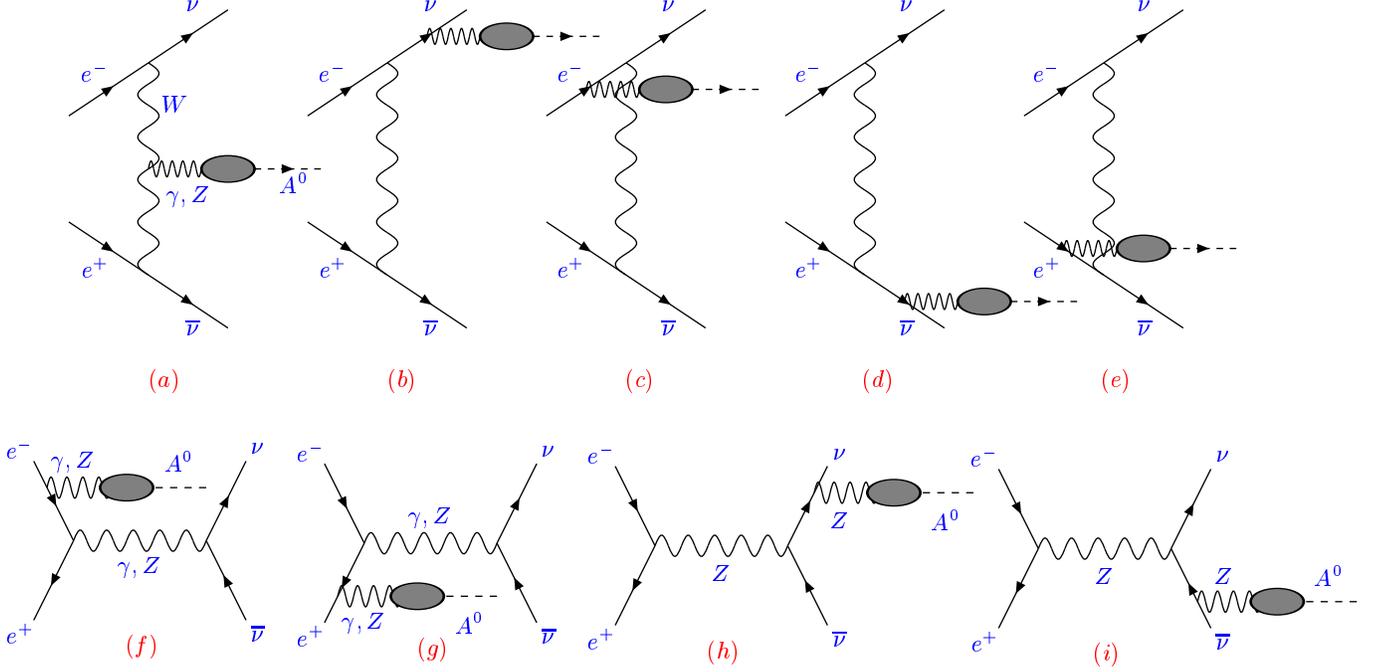}
\caption{Feynman diagrams for the contribution to $e^+e^- \to
\nu\anti{\nu}\ha$ through $Z$-$\ha$ and $\gamma$-$\ha$ mixing self-energies.
The blob denotes the renormalized $Z$-$\ha$ and $\gamma$-$\ha$ mixing. }
\label{fig:ZA}
\end{figure}

Another set of diagrams involving $Z-\ha$ mixing is given in 
Fig.~\ref{fig:ZA}.  
The contribution from diagram (a), in which the 
$\wp\wm$ couple to a virtual $Z^*$ which then mixes with the $\ha$ 
via (infinite) one-loop diagrams, is zero for on-shell
$W$ bosons \cite{gunhabkao}.  With virtual $W$ bosons as in our case, 
diagram (a) is non-zero.  However, it is not gauge invariant on its own.
Additional diagrams (b)-(e) have to be taken into account.  
The sum of all these diagrams gives zero as a consequence of gauge invariance.
This can be seen as follows.
The one-particle-irreducible (1PI) 
two-point function for $Z-\ha$ mixing is defined as 
$-ip_{A}^\mu\Sigma_{Z-\ha}(p_A^2)$, where $p_{A}^\mu$ is also the 
off-shell $Z$ momentum.  Gauge invariance for the $Z$ tells
us that after summing over all possible $Z$ attachments we must
have $p_{\ha}^\mu \cala_\mu=0$, where $\cala_\mu$ is the full
amplitude that would be dotted into the $Z$ propagator.  
In some renormalization schemes, one can also have
$\gam-\ha$ mixing, in which case a similar
argument guarantees
that after a complete sum over the diagrams (a), (c) and (e) 
with a virtual $\gam^*$ one gets zero, just as in the $Z^*$ case.  
Of course, there are renormalization schemes (such as that 
chosen in Ref.~\cite{cpr}) in which there is no $\gam-\ha$ mixing 
and this issue does not arise.
Similarly, the sum of $s$-channel diagrams (f)-(i) also gives zero. 

\begin{figure}
\includegraphics*[20,650][600,770]{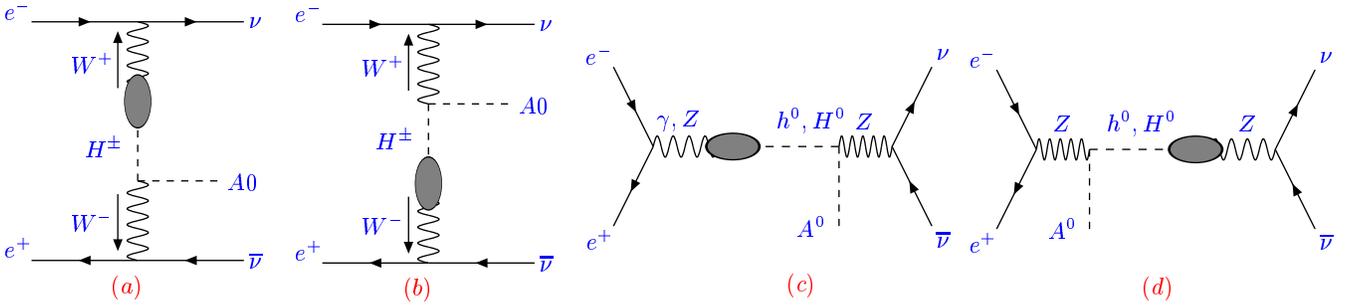}
\caption{Contributions to $e^+e^- \to \nu\anti{\nu}\ha$ that we neglect. 
The blob denotes the renormalized 
$W^{\pm}$-$H^{\pm}$ mixing in diagrams (a) and (b)
and the renormalized $(\gamma,Z)$-$(\hl,\hh)$ 
mixing in diagrams (c) and (d).}
\label{fig:WHA}
\end{figure}

The diagrams shown in Fig.~\ref{fig:WHA} do not contribute to our calculation.
For diagrams (a) and (b), 
the one-loop $\wp\to\hp$ mixing graph must be proportional to
$k_{\wp}^\mu=p_{\nu}^\mu-p_{e^-}^\mu$ (for example) , 
which gives zero when acting on the $V_{\mu}\anti{\nu}\gamma^{\mu}e$ vertex
by the equation of motion.  (Here we take the approximation that 
both electron and neutrino are massless.)
Similarly, diagrams (c) and (d) also vanish.

Let us define $p_{e^-}$, $p_{e^+}$, $p_{\nu}$, $p_{\anti\nu}$,
$k_1=p_\nu - p_{e^-}$ and $k_2=p_{\anti\nu}-p_{e^+}$ to be the 
momentum for the 
incoming electron, positron, outgoing neutrino, anti-neutrino, intermediate 
$\wp$ and $\wm$, respectively.  
The $t$-channel matrix element for $\process$ can be decomposed 
as follows (using $\xi=1$ Feynman-'t Hooft gauge):
\bea
\mathcal{M}_t&=&
\frac{g^2}{2}
{\anti v(p_{e^+},s_{e^+})\gamma_\nu P_L v(p_{\anti \nu},s_{\anti \nu})
\over k_1^2-\mw^2}
{\anti u(p_\nu,s_\nu)\gamma_\mu P_L u(p_{e^-},s_{e^-})\over k_2^2-\mw^2}
 \nonumber \\ && \times
\left[F\eps^{\alpha\beta\mu\nu}k_{1\,\alpha} k_{2\,\beta}+Gg^{\mu\nu}
+ H_1 p_{e^-}^\nu p_{e^+}^\mu + H_2 p_{\nu}^\nu p_{\anti\nu}^\mu
+ H_3 (-p_{e^-}^\nu p_{\anti\nu}^\mu) + H_4 (-p_{\nu}^\nu p_{e^+}^\mu)
\right],
\label{eq:maxele}
\eea
where $P_L=(1-\gamma_5)/2$. 
For the triangle loops [Fig.~\ref{fig:WWA} diagrams (1)-(7)],
only the combination  $k_1^\nu k_2^\mu$ appears for the $H_i$ term.  Therefore,
the terms inside the square brackets in 
Eq.~(\ref{eq:maxele}) can be simplified as 
\beq
\left[ F\eps^{\alpha\beta\mu\nu}k_{1\,\alpha}k_{2\,\beta}+
G g^{\mu\nu}+ H k_1^\nu k_2^\mu \right],
\eeq
where $H=H_1=H_2=H_3=H_4$.
Recall that the index $\mu$ ($\nu$) 
is associated with the $W^+$ ($W^-$)
going {\it into} the $e^-$ ($e^+$) line.
Analytical formulae for each Feynman diagram contribution to $F$, $G$, $H_i$ 
are summarized in Appendix \ref{app:formula_t}.
In the on-shell limit where $k_1^2=k_2^2=m_W^2$, $G$ and $H$ are both
zero \cite{gunhabkao},
which can be seen explicitly in the analytical formulas.

The $s$-channel matrix element can be written as two non-interfering pieces:
$\mathcal{M}_s=\mathcal{M}_s(e^-_L e^+_R)+\mathcal{M}_s(e^-_R e^+_L)$, where 
\begin{equation}
\mathcal{M}_s(e^-_L e^+_R) = \sum_{i\ {\rm even}} \mathcal{M}_i \mathcal{O}_i,
\qquad
\mathcal{M}_s(e^-_R e^+_L) = \sum_{i\ {\rm odd}} \mathcal{M}_i \mathcal{O}_i.
\end{equation}
The definitions of the operators $\mathcal{O}_i$ and the contribution of 
Fig.~\ref{fig:ZZA} to $\mathcal{M}_i$ are given in 
Appendix \ref{app:formula_s}.

The spin averaged matrix element squared is:
\bea
{1\over 4} {\sum}_{spins}|\calm|^2&=& {1\over 4} {\sum}_{spins}|\calm_t|^2
+{3\over 4} {\sum}_{spins}|\mathcal{M}_s(e^-_L e^+_R)|^2
+{3\over 4} {\sum}_{spins}|\mathcal{M}_s(e^-_R e^+_L)|^2\nonumber \\
&+&{1\over 4} {\sum}_{spins}\calm_t\mathcal{M}_s(e^-_L e^+_R)^*
+{\rm h.c.}, 
\label{eq:maxelesqu}
\eea
which is used in the cross section calculations. 
The factor of 3 in the second and third terms represents the sum over
the three neutrino flavors in the $s$-channel contribution.
Notice that only the matrix element $\mathcal{M}_s(e^-_L e^+_R)$ interferes 
with the $t$-channel diagrams, while the other 
$s$-channel matrix element does not.
The explicit expression for the pieces in Eq.~(\ref{eq:maxelesqu}) is 
given in Appendix~\ref{app:square}.

Our numerical computations were performed using the LoopTools
package~\cite{Hahn:1998yk}.  
We thus write the Appendices using the notation of LoopTools
\cite{Hahn:1998yk} for the one-loop integrals.

\section{\label{sec:numerics}Numerical Results}

\subsection{\label{subsec:MSSM2HDM}2HDM with tree-level MSSM 
mass and coupling relations}

In this section, we give results for the $\epem\to\nu\anti\nu \ha$
cross section as a function of $\mha$ and $\tanb$ assuming
a type-II 2HDM with the tree-level MSSM constraints on the 
quartic couplings in the Higgs potential. 
As a result, the masses $\mhl$, $\mhh$ and $\mhpm$ 
and the mixing angle $\alpha$ of the CP-even sector 
are all fixed in terms of $\mha$ and $\tan\beta$
by the tree-level relations of the MSSM.
The tree-level MSSM couplings lead to a theoretical upper bound
on the mass of the lighter CP-even $\hl$ of $m_{h^0} \leq m_Z$
(which is increased to
$\sim 135\gev$ by radiative corrections \cite{hmass}). 
In the decoupling region of large $\mha$ (typically $\mha\gsim 2\mz$
is large enough),
the only light Higgs boson is the CP-even $\hl$, whose couplings  
to the SM particles approach their SM values.  
The other Higgs bosons $\hh$, $\ha$ and $\hpm$ can be as heavy as a TeV.  
The heavy Higgs bosons are nearly degenerate in mass, with mass 
splittings of the order of $\mz^2/\mha$. 

Our choice of the tree-level MSSM Higgs sector can be thought of
simply as a representative model choice within the 2HDM
that gives decoupling and a SM-like $\hl$ as $\mha$ gets large.
It will allow us to explain general features of the cross
section and how they depend upon $\mha$ and $\tanb$.
We will only examine results for $\tanb>1$ in this section.
Our focus will be on situations in which $\mha\geq E_{cm}/2$,
implying that pair production of the $\ha$ (e.g., $\epem\to \nu\anti\nu\ha\ha$)
will be kinematically forbidden, 
as will $\epem\to\ha\hh$, since $m_{H^0} \simeq \mha$ 
for MSSM-like mass relations.
The $\epem\to\hl\ha$ process will also
be strongly suppressed for $\mha\gsim 2\mz$.  Thus, we are considering
situations in which $\ha$ discovery might only be possible via
the (one-loop) single $\ha$ production mechanisms that we consider.

\begin{figure}
\resizebox{0.8\textwidth}{!}{
\rotatebox{90}{\includegraphics{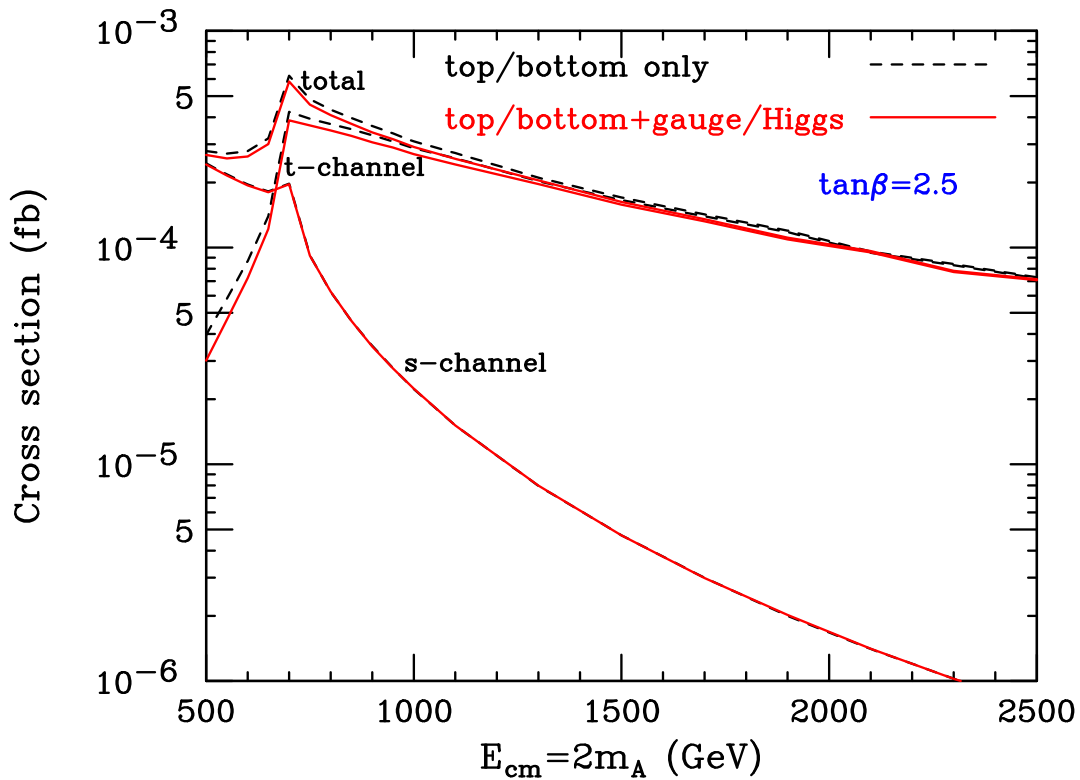}}}
\caption{Behavior of the $t$- and $s$-channel contributions to
$e^+e^- \to \nu \anti \nu \ha$ with $E_{cm}$.  Here $\tan\beta = 2.5$
and $\mha = E_{cm}/2$. The line labeled ``total'' shows the sum of the $t$-
and $s$-channel contributions including their interference.
The $s$-channel results including gauge/Higgs contributions
are indistinguishable from the $s$-channel results with only
top/bottom loops.}
\label{fig:nnArootS}
\end{figure}

The first feature of interest is that
the $s$-channel and $t$-channel contributions to the cross section
have different behavior as the collider center-of-mass energy increases,
as shown in Fig.~\ref{fig:nnArootS}. 
In particular, for $\mha = E_{cm}/2$, the $s$-channel contribution dominates
for $E_{cm} \lsim 650$ GeV, while the $t$-channel contribution dominates
for $E_{cm} \gsim 700$ GeV.
To some extent the $s$- and $t$-channel contributions can be
separated experimentally.  For example, the $s$-channel contributions
can be isolated by looking at $Z\ha$ final states in which
the $Z$ decays to an observable final state, such as $q\anti q$
or $\ell^+\ell^-$.  The $t$-channel contributions can be isolated
to a large extent by looking at the $\ha$ in a visible final state
decay mode (for example, $b\anti b$ or $\tau^+\tau^-$), reconstructing
the mass recoiling against the $\ha$, and demanding that this
recoil mass not be close to $\mz$.  In this latter case, this selection
procedure would reduce somewhat the $t$-channel cross sections presented
(which are integrated over all recoil masses).

\begin{figure}
\resizebox{\textwidth}{!}{
\rotatebox{90}{\includegraphics{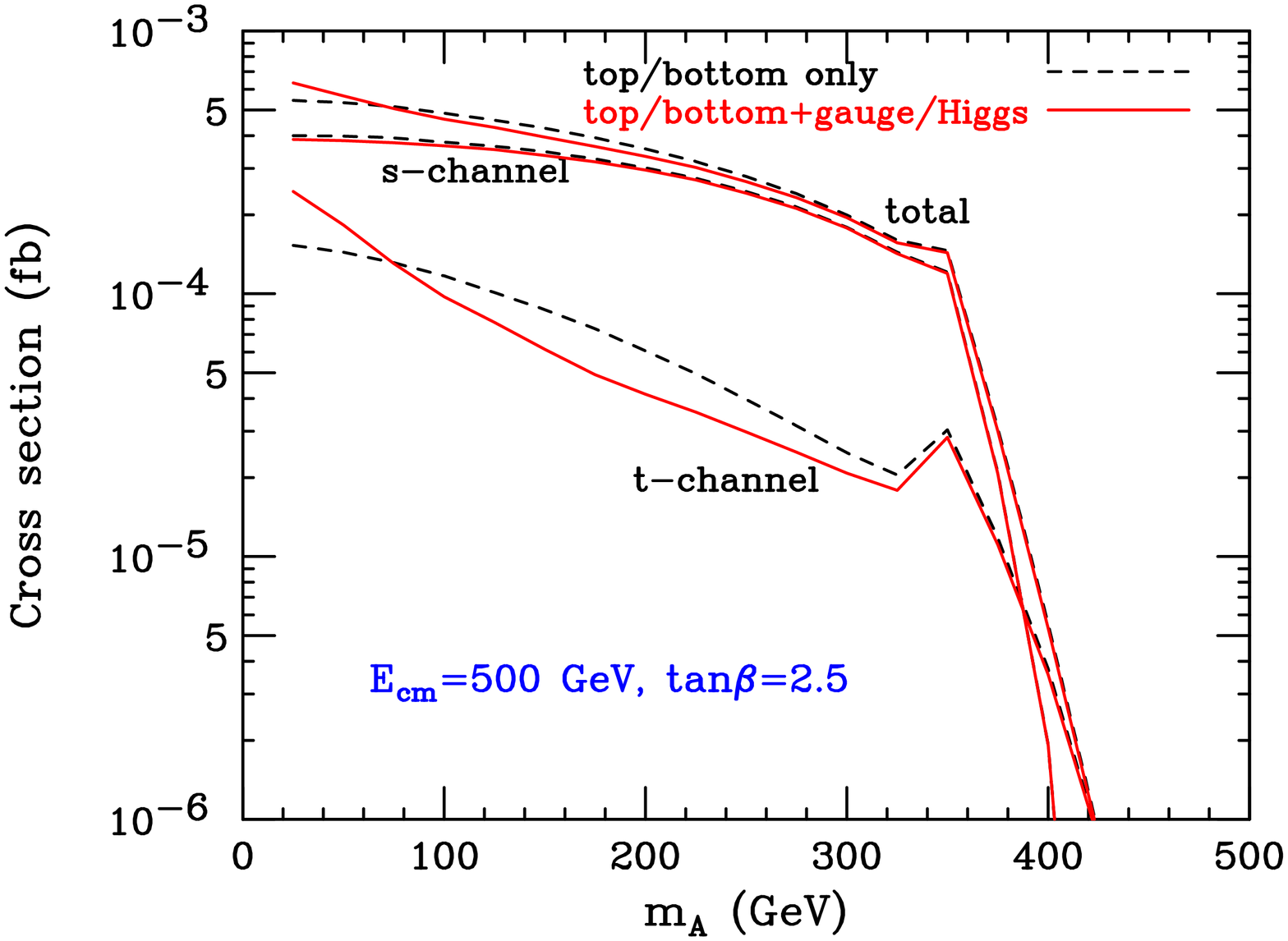}}
\rotatebox{90}{\includegraphics{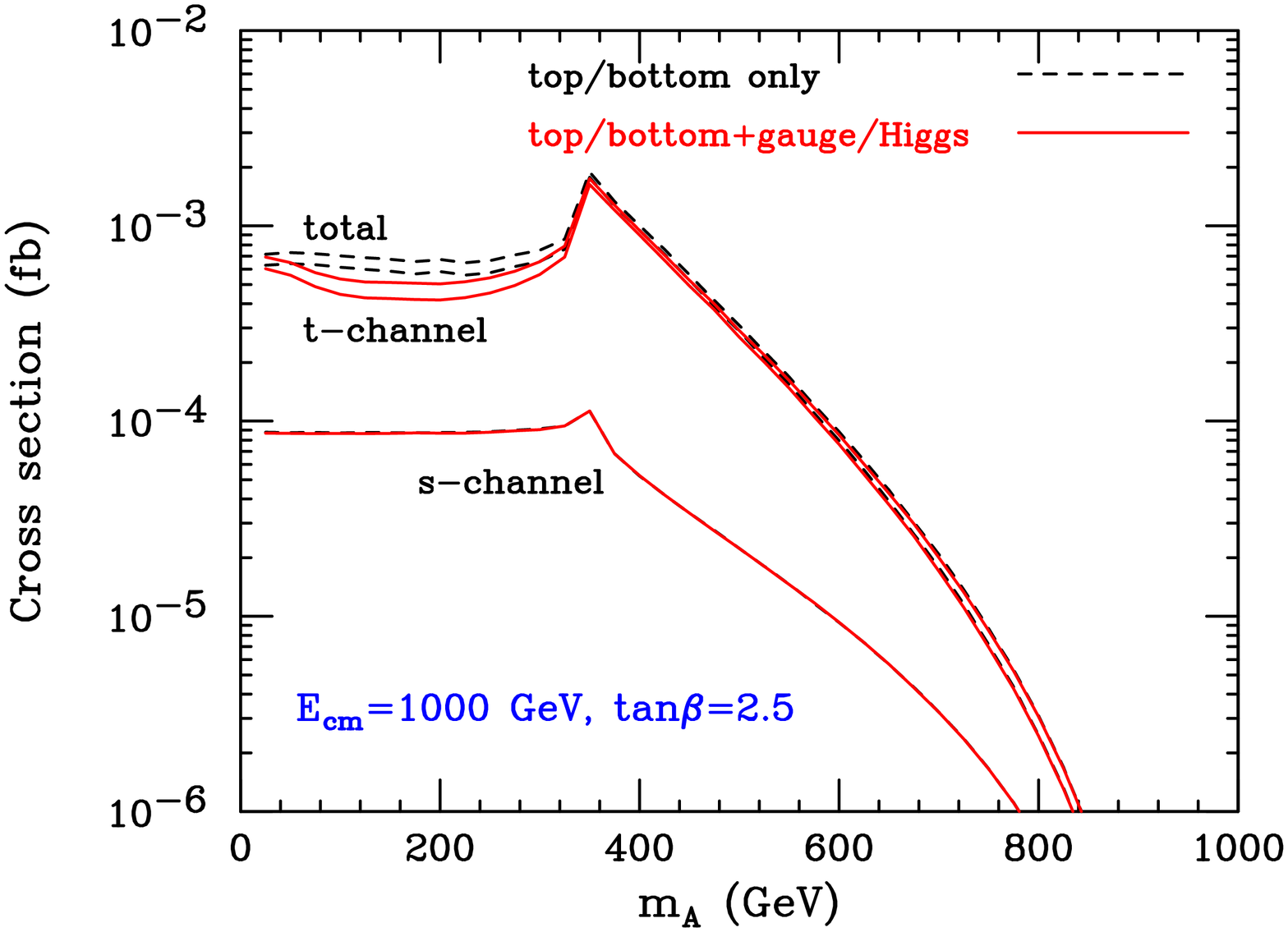}}}
\caption{Cross section for $e^+e^- \to \nu \anti \nu \ha$ as a function
of $\mha$, for $\tan\beta = 2.5$ and $E_{cm} = 500$ GeV (left) and 
1000 GeV (right).  The line labeled ``total'' shows the sum of the $t$-
and $s$-channel contributions including their interference.
 As in the previous figure,
the $s$-channel top/bottom only and top/bottom+gauge/Higgs lines
are essentially identical.
In the $E_{\rm cm}=1000\gev$ plot, 
we note that the ``$t$-channel'' lines (whether top/bottom only
or top/bottom+gauge/Higgs) are always below the ``total'' lines
in the $\mha\sim 200\gev$ region. Thus, for instance,
the upper dashed line is the total ``top/bottom only'' result
while the lower dashed line is the $t$-channel ``top/bottom only''
result. }
\label{fig:nnAmA}
\end{figure}

Perhaps most importantly, we find that
the cross sections are quite small, generally below 0.001 fb.
This holds even for rather low values of $\mha$.
The cross section as a function of $\mha$ is shown in Fig.~\ref{fig:nnAmA}.
This figure shows the expected peak in $\sig(\nu\anti\nu\ha)$ 
at $\mha\sim2\mt$ from the top quarks in the loop going on shell,
followed by a rapid fall as one approaches the kinematic limit
at $\mha=\sqrt s$. 
Notice that for $E_{cm} = 500$ GeV, the $s$-channel contribution dominates,
while for $E_{cm} = 1000$ GeV, the $t$-channel contribution dominates,
as expected from Fig.~\ref{fig:nnArootS}.
Similar results were presented in Ref.~\cite{Arhrib:2002ti}.
While we include all three flavors
of neutrinos in the final state since they are experimentally 
indistinguishable, Ref.~\cite{Arhrib:2002ti} included only 
$\nu_e \bar \nu_e$ in the final state, leading to an $s$-channel
cross section smaller by a factor of 3 than our result.
Taking this into account, our results are in rough agreement 
with those of Ref.~\cite{Arhrib:2002ti}.
For example, for the point $E_{cm} = 500$ GeV, $m_A = 100$ GeV,
$\tan\beta = 2.5$, and one neutrino flavor, our result is about a factor
of two larger than that shown in Fig.~4 of Ref.~\cite{Arhrib:2002ti}.
Part of the discrepancy, a factor of $(137/128)^4 = 1.3$, is explained
by the use of $\alpha = 1/137$ in Ref.~\cite{Arhrib:2002ti} versus
our use of $\alpha = 1/128$.\footnote{This was pointed out
in a private communication
from the author of \cite{Arhrib:2002ti} in which he also gives numbers
(that include the $\alpha$ correction) that are in close agreement with ours.}

\begin{figure}[h!]
\resizebox{\textwidth}{!}{
\rotatebox{90}{\includegraphics{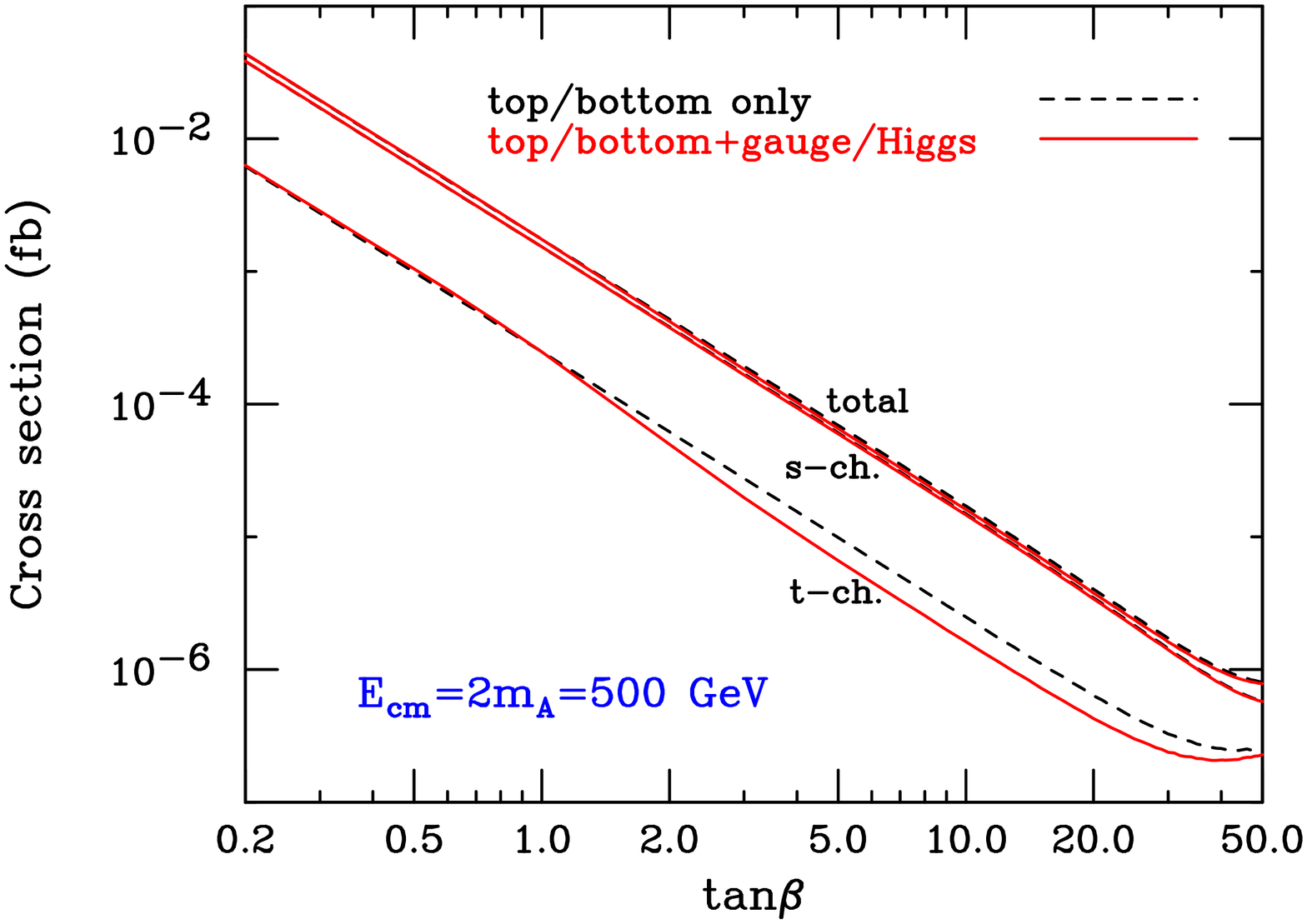}}
\rotatebox{90}{\includegraphics{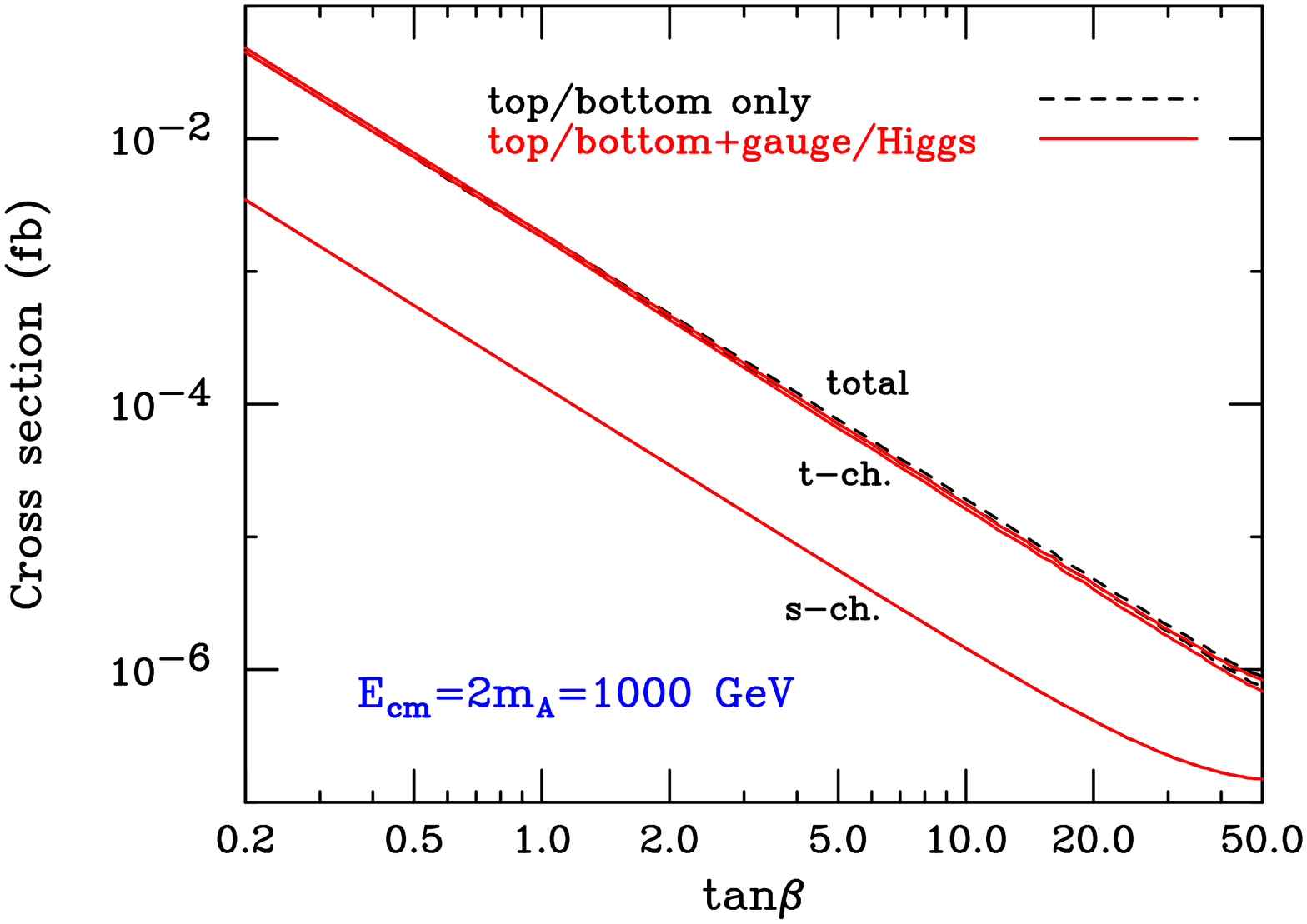}}}
\caption{$\tan\beta$ dependence of the cross section 
for $e^+e^- \to \nu \anti \nu \ha$, for $E_{cm} = 2 \mha = 500$ GeV (left)
and 1000 GeV (right). As in the previous two figures, the $s$-channel
curves with and without the gauge/Higgs contributions are
indistinguishable. In the $E_{\rm cm}=500\gev$ ($E_{\rm cm}=1000\gev$) case, 
the ``total'' curves are slightly higher
than the ``$s$-ch.'' (``$t$-ch.'') 
curves for both ``top/bottom only'' and for
``top/bottom+gauge/Higgs''. The ``top/bottom only'' curves are slightly
above the ``top/bottom+gauge/Higgs'' curves in both the ``$s$-ch.''
(``$t$-ch.'') and ``total'' cases. In this figure, we have
used the shorthand ``ch.'' for ``channel''.}
\label{fig:nnAtanb}
\end{figure}

The dependence of the cross section on $\tan\beta$ is shown in
Fig.~\ref{fig:nnAtanb}. 
This plot clearly
shows that for the MSSM-like parameter relations, 
detection of the $\nu\anti\nu\ha$ final state will only be possible if
$\tanb\lsim 1$. 
The cross section falls like a power law with increasing $\tan\beta$.
This is due to the fermion triangle diagrams
with a top quark coupling to $\ha$ (diagram (1) in Fig.~\ref{fig:WWA}
and the $t$-loop case for diagrams (1) and (2) in Fig.~\ref{fig:ZZA}),
which dominate at low $\tan\beta$ and give a cross section 
proportional to $y_t^2 \sim (\tan\beta)^{-2}$.  
At large values of $\tan\beta \gsim 30$,
the fermion triangle diagrams with a bottom quark coupling to $\ha$
(diagram (2) in Fig.~\ref{fig:WWA} and the $b$-loop
case for diagrams (1) and (2) in Fig.~\ref{fig:ZZA})
begin to contribute significantly and affect the dependence on $\tan\beta$, 
since these diagrams give a cross section proportional to 
$y_b^2 \sim (\tan\beta)^2$.

\begin{figure}[h!]
\begin{center}
\includegraphics[width=6.2cm,height=8.7cm,angle=90]{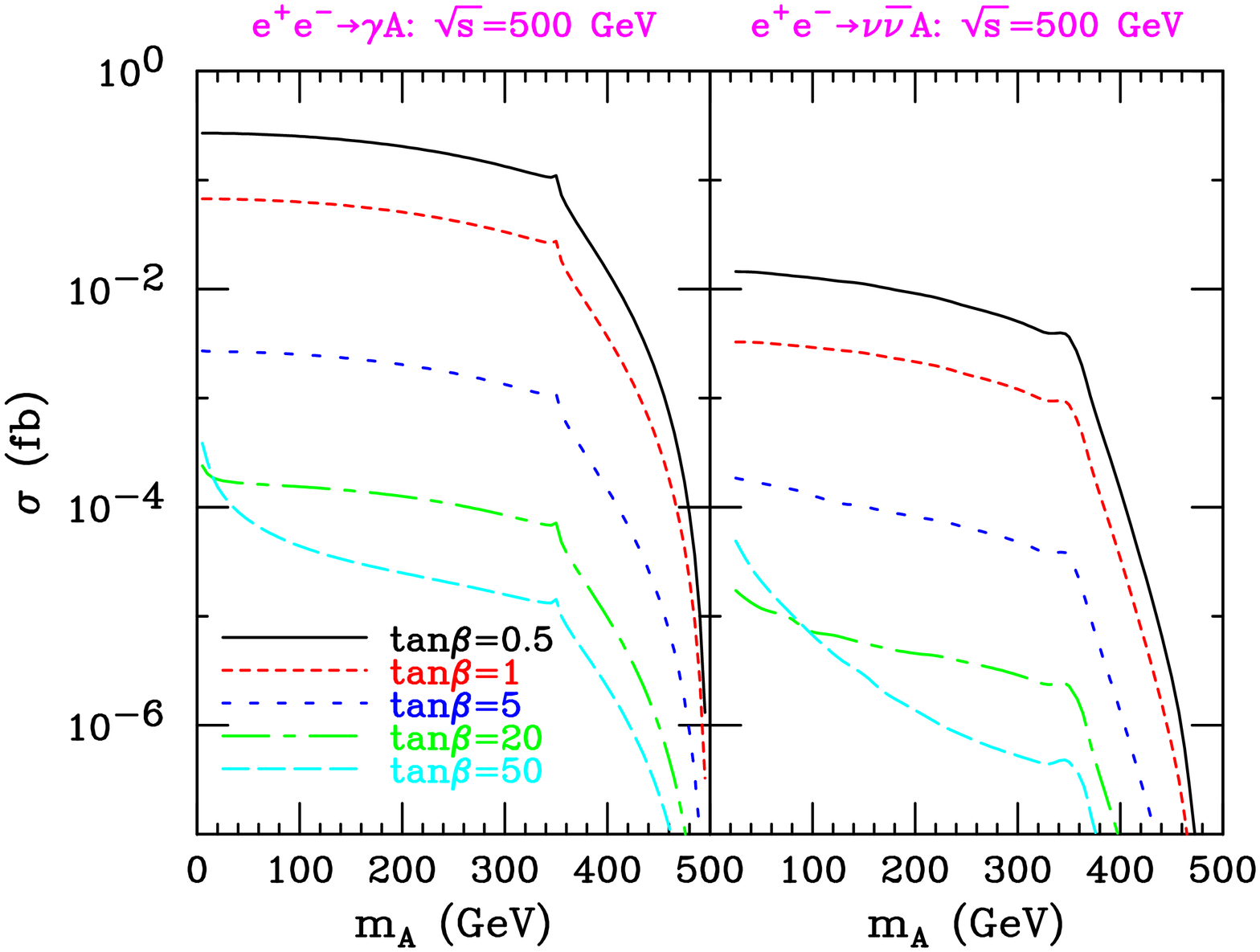}
\includegraphics[width=6.2cm,height=8.7cm,angle=90]{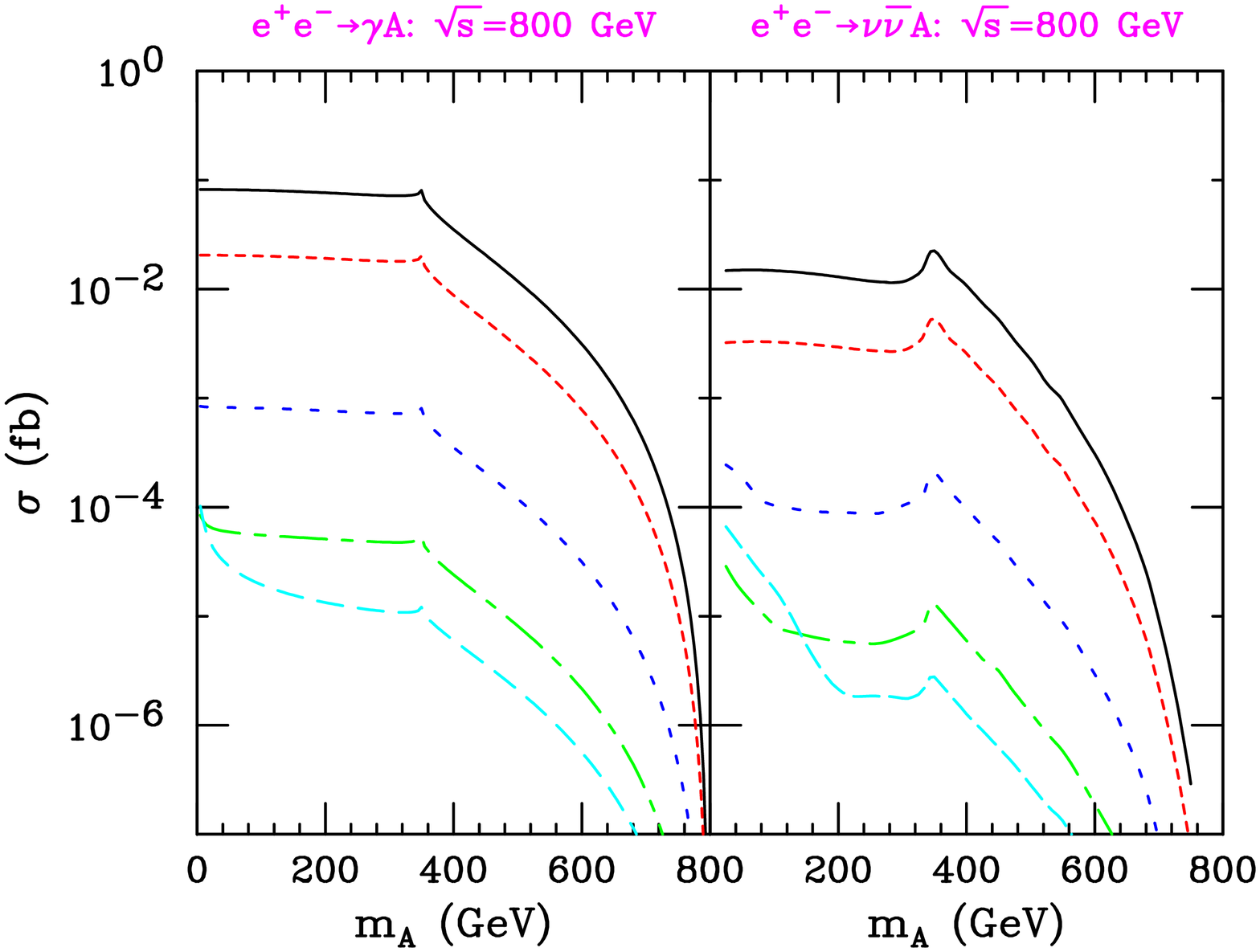}
\caption{\label{loopha} The $\epem\to\gam\ha$ and $\nu\anti\nu\ha$ 
cross sections as a function of $\mha$ for $\rts=E_{\rm cm}=500\gev$ 
and $800\gev$,
for $\tanb=0.5,1,5,20,50$. We employ the tree-level MSSM parameterization.
The $\epem\to \nu\anti\nu\ha$ cross section includes all contributions.}
\end{center}
\end{figure}

Another perspective on these results, and a comparison to the $\epem\to\gam\ha$
process~\cite{Djouadi:1996ws,Akeroyd:1999gu} 
is presented in Fig.~\ref{loopha}. 
For the tree-level MSSM type of 2HDM parameter choices,
the $\epem\to\gam\ha$ process would probably lead to earlier
discovery of the $\ha$ than would the $\epem\to \nu\anti\nu\ha$
assuming both have small background.  However, as described
in the next section, the $\epem\to\nu\anti\nu\ha$ can
be enhanced for non-decoupling 2HDM parameter choices that
lead to large Higgs self-couplings, whereas the $\epem\to\gam\ha$
cross section is not sensitive to Higgs self-couplings and would
not be enhanced in such a parameter regime. 
Note also that the advantage
of the $\epem\to\gam\ha$ process over the $\epem
\to\nu\anti\nu\ha$ process  decreases slowly with increasing $E_{\rm cm}$,
as seen from the figure by comparing the results for
$E_{\rm cm}=500\gev$ to those for $E_{\rm cm}=800\gev$.

\begin{figure}[h!]
\resizebox{\textwidth}{!}{
\rotatebox{0}{\includegraphics{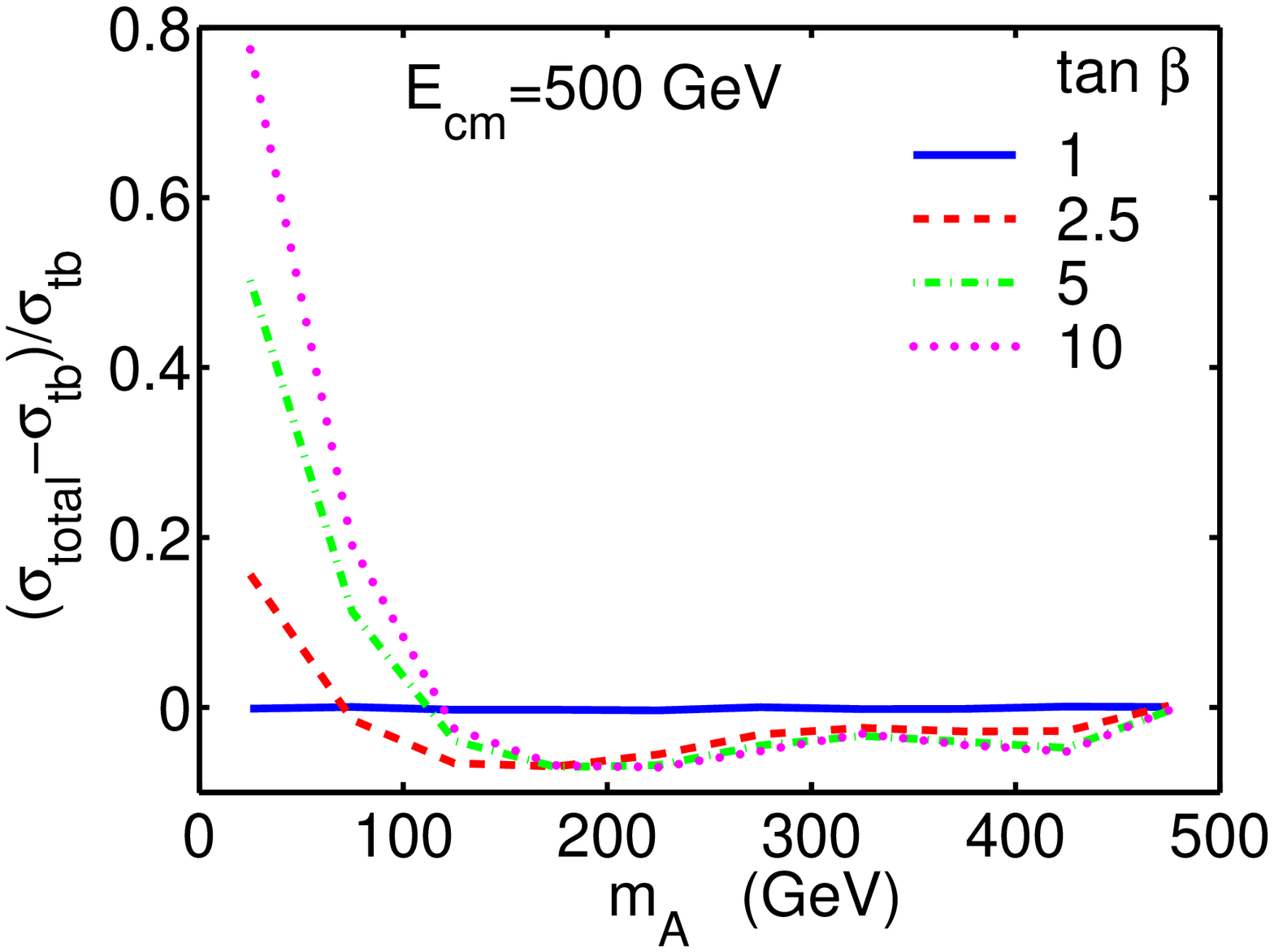}}
\rotatebox{0}{\includegraphics{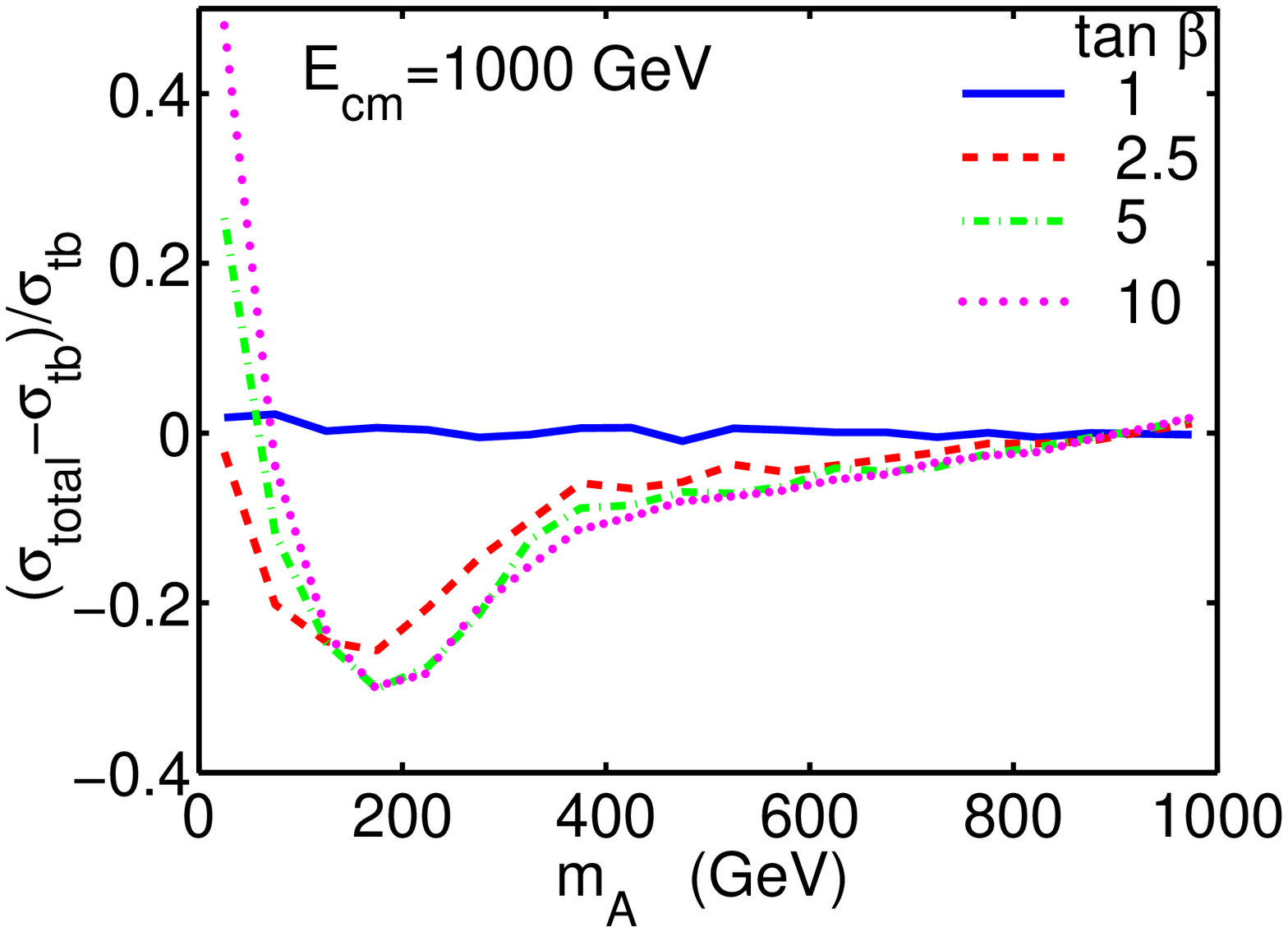}}}
\caption{Contribution of the gauge and Higgs boson loops relative to the
top and bottom quark loops, as a function of $\mha$.
Plotted is the gauge and Higgs contribution normalized
to the top and bottom quark contribution, 
$(\sigma_{total} - \sigma_{tb})/\sigma_{tb}$.
}
\label{fig:nnAcorr}
\end{figure}

The relative size of the gauge and Higgs boson contributions compared
to the top and bottom quark contributions is illustrated in 
Fig.~\ref{fig:nnAcorr}.
For $\tan\beta = 1$, the gauge and Higgs boson loops vanish.  For larger
values of $\tan\beta$, the contribution of the gauge and Higgs boson loops
relative to that of the top and bottom quark loops typically increases
with increasing $\tan\beta$.
At $E_{cm} = 500$ GeV, the gauge and Higgs boson contributions can be 
quite significant at low $\mha \lsim 100$ GeV, especially for larger
values of $\tan\beta$.  However, in the MSSM, $\mha \lsim 92$ GeV is 
excluded by the LEP II data~\cite{lepa,lep2}.  
For larger values of $\mha$, the 
gauge and Higgs loops interfere destructively with the dominant top 
and bottom quark loops, resulting in a reduction of the cross section 
by less than 10\%.  
At $E_{cm} = 1000$ GeV, in contrast,
the destructive interference of the gauge and Higgs
boson loops with the top and bottom quark loops is much more significant
for $\mha$ below the top quark pair production threshold of 350 GeV,
suppressing the cross section by as much as 30\% for $\mha \sim 200$ GeV.
For $\mha \gsim 350$ GeV, the gauge and Higgs boson loops reduce
the cross section by less than 10\%.  
The change in the relative size of the gauge and Higgs boson loops 
at different center-of-mass energies can be understood as follows.
At $E_{cm} = 500$ GeV, the cross section
is dominated by the $s$-channel diagrams.  The gauge and Higgs boson
contributions to the $s$-channel matrix element come only 
from box diagrams [diagrams (3) and (4) in Fig.~\ref{fig:ZZA}].
At $E_{cm} = 1000$ GeV,
the cross section is dominated by the $t$-channel diagrams.  The gauge 
and Higgs boson contributions to the $t$-channel matrix element come
from both triangle diagrams and box diagrams 
[diagrams (3)-(7) and (8)-(11), respectively, in Fig.~\ref{fig:WWA}].  
The gauge and Higgs boson triangle diagrams in general give larger 
contributions to the cross section than the box diagrams,
leading to larger gauge and Higgs boson contributions 
to the $t$-channel process than to the $s$-channel process.
This behavior can also be seen in Fig.~\ref{fig:nnAmA}.

\subsection{\label{subsec:2HDM}General 2HDM}

There is, of course, much more freedom in the general 2HDM
than we have allowed for in the previous section 
First, there is the possibility of allowing for type-I fermionic couplings
as opposed to the type-II fermionic couplings employed so far.
As shown in Eq.~(\ref{typei}), the type-I $t\anti t$  coupling
is the same as the type-II coupling.  This implies that the dominant
$t$ loop contribution to the $\nu\anti\nu\ha$ cross section will
be unchanged.  The type-I $b\anti b$ coupling is proportional to $\cotb$
as opposed to $\tanb$ for type-II coupling; this means
that the $b$-loop contribution to the $\nu\anti\nu \ha$ cross
section is never important for type-I couplings. 
Numerically, this implies that
the leveling off of $\sig(\nu\anti\nu\ha)$ at $\tanb\gsim 40$
in Fig.~\ref{fig:nnAtanb} (and eventual rise at still
larger $\tanb$) would not take place for type-I couplings.
There is also the possibility of so-called type-III fermion couplings in which
the two Higgs doublets both couple to up and down type quarks.
In general, such couplings yield flavor changing neutral currents
that are too large compared to existing experimental constraints.
In addition, the numerical modifications to the type-II predictions
already given would not be large.  Thus, we do not consider type-I or type-III
couplings further.

A second variation in the general 2HDM context is to allow for
CP-violating couplings.  As explained
in the introduction, we have chosen not to explore this possibility
here as it makes the $\ha$ less unique and because
considerable cancellations between CP-violating contributions
deriving from the Higgs sector are required for consistency between
the computed EDMs and
$(g_\mu-2)$ and experimental data.  
If the Higgs sector is CP-violating, all the neutral
Higgs bosons mix and will all have some level of $VV$ coupling.
In most scenarios, all three of the neutral Higgs
bosons $h_i$ ($i=1,2,3$) would be easily detected in
$Zh_i$ production and the tree-level contributions
to the $\nu\anti\nu h_i$ ($i=1,2,3$) cross sections would all
be considerably larger than the one-loop  
contributions~\cite{Akeroyd:2001kt}.

The most interesting issue in the general CP-conserving 2HDM context
is the extent to which the Higgs self-couplings could deviate
from those in our previous MSSM-like analysis,
so that $\sig(\nu\anti\nu\ha)$ might be
substantially increased or decreased by the Higgs boson loops
compared to the value obtained from the $t$ and $b$ loops. 
In Fig.~\ref{fig:nnAcorr} we found that with MSSM-like couplings,
the gauge and Higgs boson loops could change the cross section 
by as much as 70\% compared to the $t$ and $b$ quark contributions
for low values of $\mha$, or by up to 30\% for $\mha \gsim m_Z$.
Here, we explore the effect of removing the MSSM constraint on the 
Higgs boson self-couplings $\lam_i$, while still requiring that
they remain perturbative.
(Following \cite{Gunion:2002zf}, we define
perturbativity by the requirement $\lam_i/(4\pi)\lsim \calo(1)$.)
Another constraint on the 2HDM parameters derives
from precision electroweak data, as conveniently summarized by 
the $S$ and $T$ parameters.
We will explore the extent to which
the $\nu\anti\nu \ha$ cross section can be enhanced 
via the triangle graphs involving Higgs self-couplings while
remaining consistent with the perturbativity and $S,T$ constraints.
In the preceding section, we considered parameter choices that correspond
to rapid decoupling as $\mha$ increases beyond $\mathcal O(\mz)$.
The Higgs self-coupling effects are likely to be most significant
in those regions of parameter space that are far from the decoupling
limit. 

The triangle diagrams involving one or more internal gauge bosons,
Fig.~\ref{fig:WWA} diagrams (4)-(7), 
are controlled by $VSS$ and $VVS$ couplings which are
determined by gauge invariance.  The size of these couplings is limited,
as shown in Tables~\ref{tab:coups45} and \ref{tab:coups67}.
Any significant enhancement must come from the purely Higgs loop graph
of Fig.~\ref{fig:WWA} diagram (3), which is controlled by
the Higgs self-couplings given in Table~\ref{tab:coups3}.
For the diagrams in Fig.~\ref{fig:WWA} diagram (3) 
involving one or more Goldstone bosons, the couplings are
determined by gauge invariance purely in terms of the Higgs masses, the
angle combination $\beta-\alpha$ and the weak mixing angle $\theta_W$
(see Table~\ref{tab:coups3}). 
For the diagrams with $h^0H^+\ha$ or $H^0H^+\ha$ running in the loop,
on the other hand, the couplings $g_{\ha\ha\hl}$
and $g_{\ha\ha\hh}$ are determined by the
invariant combinations of the $\lam_i$, denoted by $\lamT$ and $\lamU$
in Ref.~\cite{Gunion:2002zf}, 
given in Eqs.~(\ref{lamtdef}) and (\ref{lamudef}).  These couplings
are free to vary in the general 2HDM once the MSSM constraints are
removed.
We will employ the ratios $\lamU/(4\pi)$ and $\lamT/(4\pi)$ to quantify
the perturbativity of the self-couplings. 
Numbers much larger than 1 for these ratios imply that 
the self-couplings are becoming non-perturbative and that
higher order corrections to the results we obtain could be large.

Our procedure will be to choose values for the Higgs masses
$\mha$, $\mhl$ and $\mhh$ and for $\tanb$, and then
scan over $\alpha$ and $\mhpm$ selecting points that are consistent
with the experimental $S,T$ values at the 95\% CL.
As in Ref.~\cite{Chankowski:2000an} (see also \cite{Gunion:2002zf}), 
for simplicity we
will restrict $\lam_4$ relative to $\lam_5$ by requiring
$\lam_4=-\lam_5$, implying $\lam_5v^2=(\mhpm^2-\mha^2)$.  This
choice makes it relatively easy to find
values for the other parameters that give
good agreement with precision electroweak $S,T$ data. Indeed,
as discussed
in Ref.~\cite{Chankowski:2000an}, even if we choose large values
for $\mhl$ and $\mhh$ (in particular, beyond the kinematic reach
of the linear collider), it is nonetheless possible to choose
$\alpha$ and $\mhpm$ in such a way that the $S,T$ values are within
the 95\% CL ellipse.  The key is to have $\mhpm>\mhl,\mhh$ by
a small amount (typically 10 to 30 GeV) in such a way that the
large negative $\Delta T$ generated by the heavy neutral
scalar(s) with substantial $VV$ coupling is compensated by an even larger
positive $\Delta T$ contribution coming from the $\mhpm-\mhl$
and/or $\mhpm-\mhh$ mass difference.  
As one varies $\alpha$
at fixed $\tan\beta$, it is generally possible to adjust the value
of $\mhpm$ in such a way as to remain within the (upper right hand
segment of the) 95\% CL $S$-$T$ ellipse.

\begin{figure}[p!]
\resizebox{0.9\textwidth}{!}{
\includegraphics[width=3in,height=7in,angle=90]{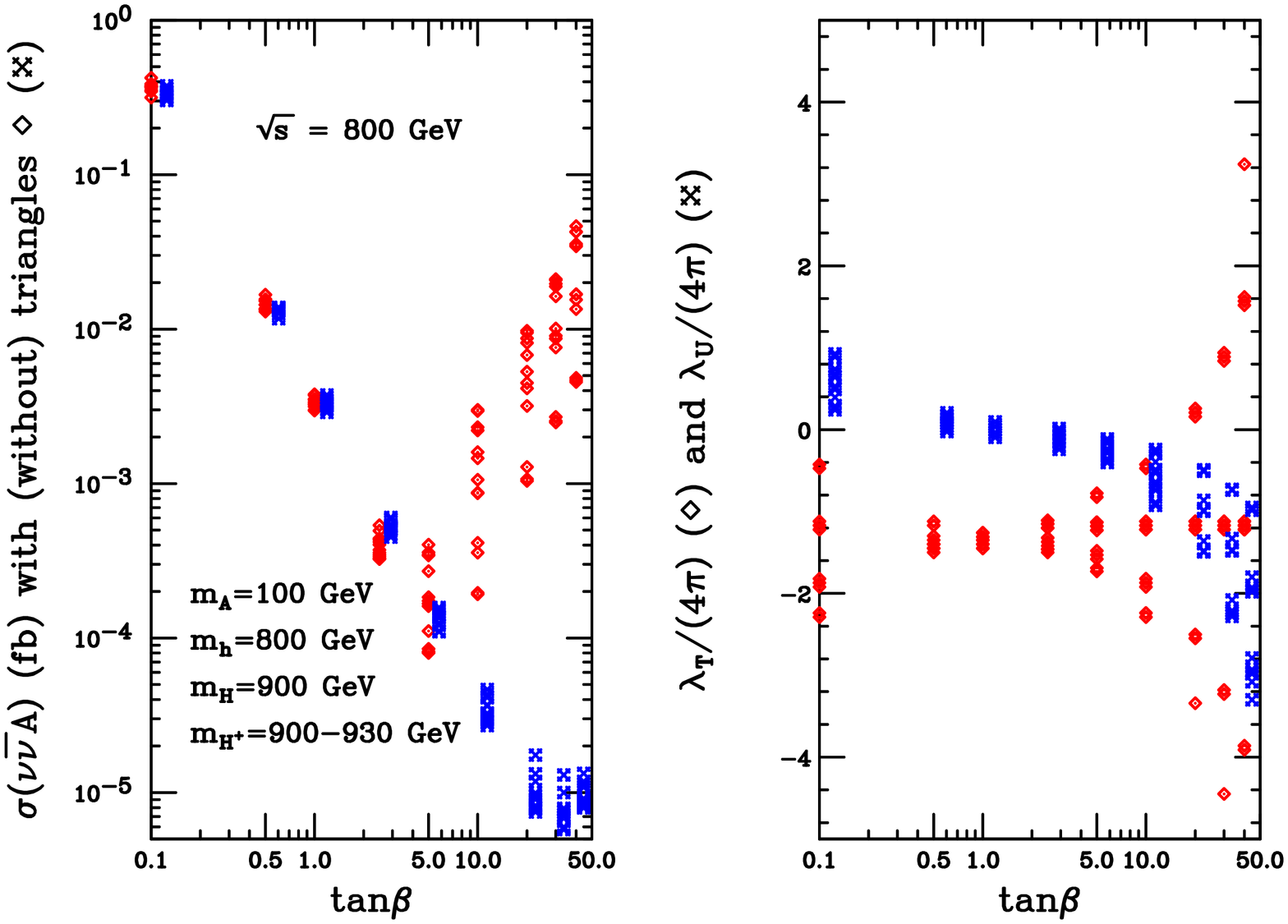}}
\resizebox{0.9\textwidth}{!}{
\includegraphics[width=3in,height=7in,angle=90]{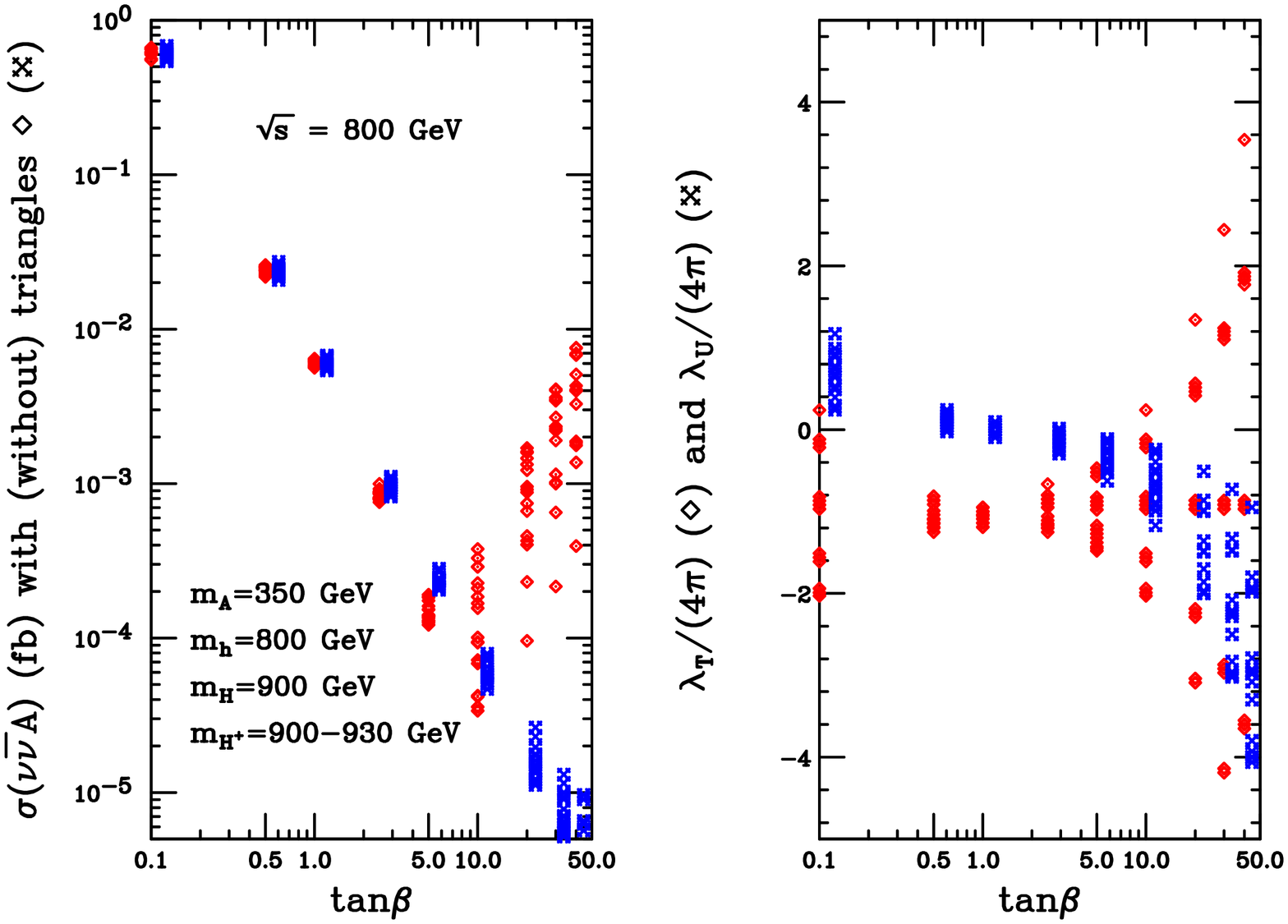}}
\resizebox{0.9\textwidth}{!}{
\includegraphics[width=3in,height=7in,angle=90]{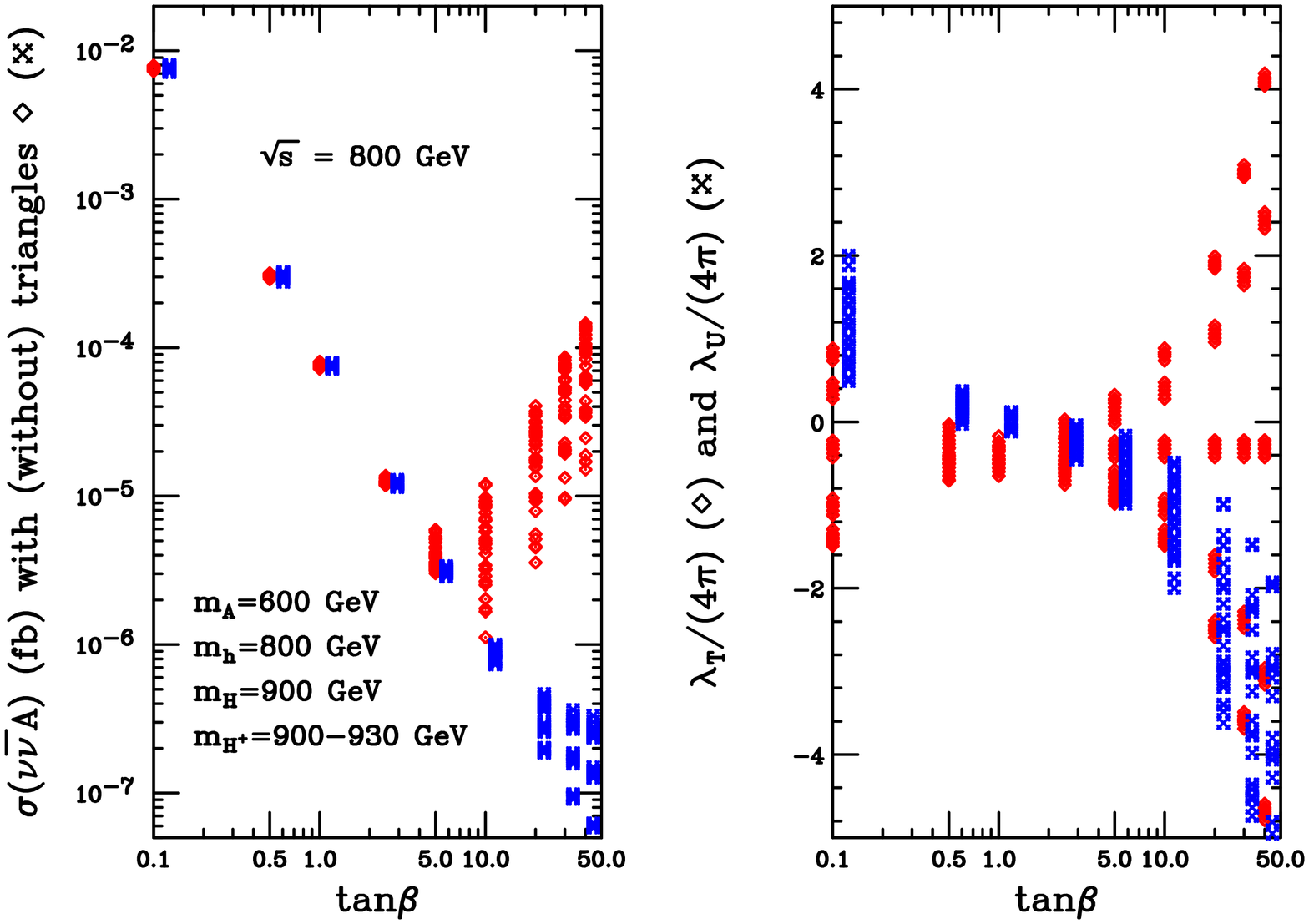}}
\caption{Left-hand graphs: 
$\sig(\nu\anti\nu \ha)$ as a function of $\tan\beta$ 
for $E_{cm} = 800$ GeV, with (points slightly to the left) and without
(points slightly to the right) Higgs-self-coupling
triangle diagrams. Right-hand graphs: corresponding values
of $\lam_{T,U}/(4\pi)$. (We do not attempt
to display point-to-point matching between the left-hand
and right-hand plots.) The three plots are for $\mha=100,350,600\gev$,
$\mhl=800\gev$, $\mhh=900\gev$ and $\mhpm\in[900,930]\gev$.
The plotted points have $\mhpm$ and $\alpha$ chosen so that
$S,T$ lie within the 95\% CL precision electroweak data ellipse.}
\label{c2hdm}
\end{figure}

We show the results of this procedure in Fig.~\ref{c2hdm}.
We compute $\sig(\nu\anti\nu\ha)$ for
$\mha=100$, $350$, and $600\gev$ as a function of $\tanb$, 
assuming a collider energy of $\rts=800\gev$.
We take $\mhl=800\gev$ and $\mhh=900\gev$ and scan
over $0\leq\alpha\leq \pi$ in 10 steps and over $\mhpm$ in steps
of $10\gev$ beginning with $\mhh$ as the lowest value.
In particular, all the Higgs masses other than $\mha$
are chosen such that none of the other Higgs bosons can be produced
for the assumed collider energy.
We plot only those points for which 
the $S,T$ values are within the 95\% CL precision
electroweak ellipse. In the left-hand plots of Fig.~\ref{c2hdm} 
we give $\sig(\nu\anti\nu\ha)$ with and without including the Feynman
diagrams containing Higgs self-couplings. 
In the right-hand plots of Fig.~\ref{c2hdm},
we plot the corresponding values of $\lamT/(4\pi)$ and $\lamU/(4\pi)$
(without attempting a point-by-point identification).
We observe that the Higgs self-coupling diagrams can have a very large
effect on the cross section 
at large $\tanb$ if one is willing to accept values of $\lam_{T,U}/(4\pi)$
of order 2 to 3.  
At large $\tanb$, the  cross section is very substantially
enhanced by the self-coupling graphs and might be visible with
$L=1000\fbi$ of integrated luminosity. Certainly, detection of
this cross section would be an extremely interesting and important
probe of the Higgs self-couplings, especially given that 
all Higgs bosons other than the $\ha$ are too heavy to 
observe directly at the linear collider in the situations considered.

In short, the lesson of this section is that if nature chooses
the Higgs sector parameters to be far from the decoupling regime,
it could happen that only the $\ha$ will be within the kinematic
reach of the linear collider and that the $\nu\anti\nu \ha$
cross section might be observable.  If in the future
the LHC finds a fairly heavy CP-even Higgs boson, then, within
the 2HDM (or similar) context, the
type of situation considered here will be required for consistency
with current precision electroweak constraints and one should urgently
search for the $\ha$ in single production modes.

\subsection{\label{subsec:MSSMspecial}Special situations in the MSSM}

In order to obtain a large cross section for $\nu\anti\nu\ha$ production
in the MSSM when $\tanb>1$, the $t\anti t$ or $b\anti b$
coupling of the $\ha$ must be enhanced very substantially.  This
{\it is} within the realm of possibility. In particular, at large
$\tanb$ it is possible to have important one loop modifications to
the $\ha b\anti b$ coupling due to radiative corrections involving
a gluino and a bottom squark. We briefly review
this aspect of the MSSM and then show the resulting enhancement
for favorable parameter choices.

\begin{figure}
\includegraphics[width=4in,height=4in]{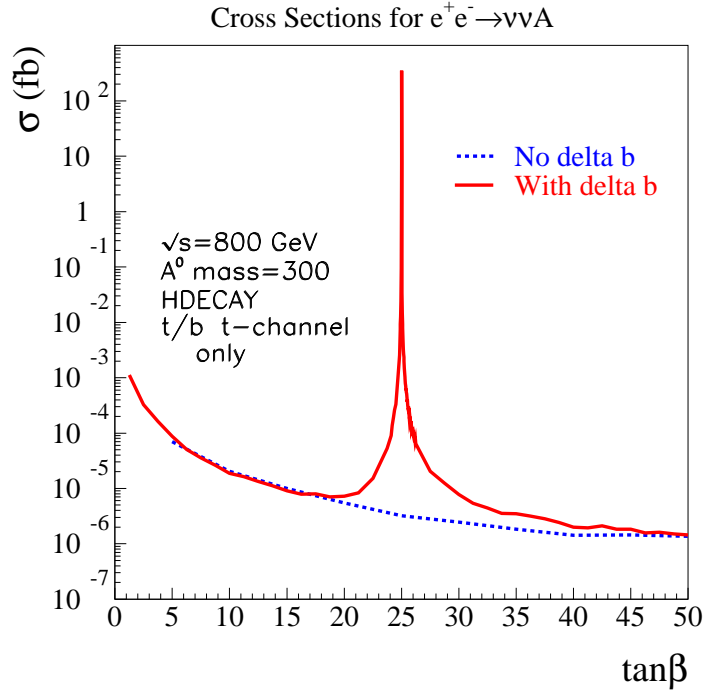}
\caption{Dependence of $\sig(\nu\anti\nu\ha)$ on $\tanb$ 
for $\mha=300\gev$ and $\rts=800\gev$, using HDECAY
and MSSM parameters as specified in the text. Results are
shown keeping only the (dominant) $t$-channel (i.e., $WW$ fusion)
$t$ and $b$ fermion-loop graphs.} 
\label{deltabe}
\end{figure}

\begin{figure}
\resizebox{\textwidth}{!}{
\includegraphics{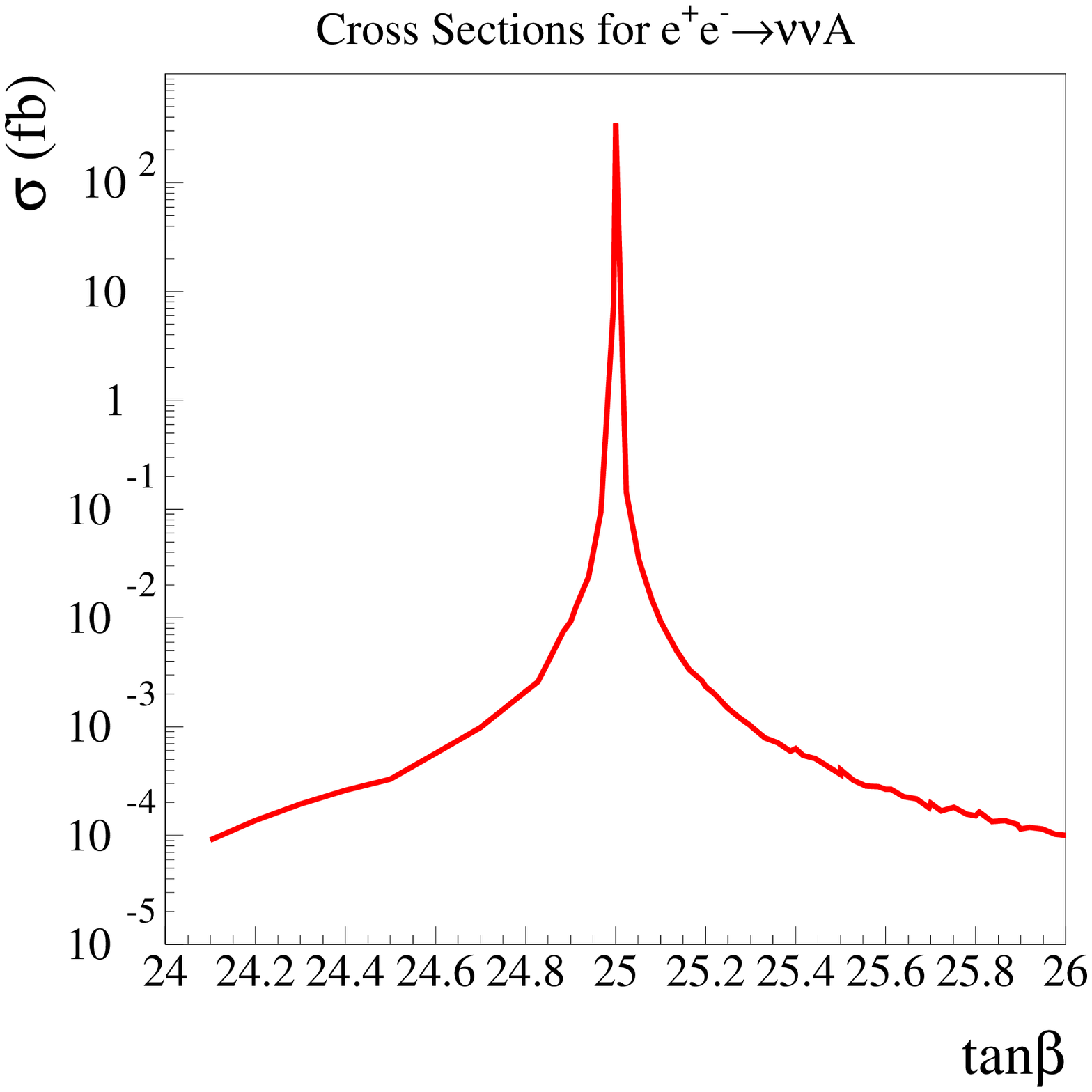}
\includegraphics{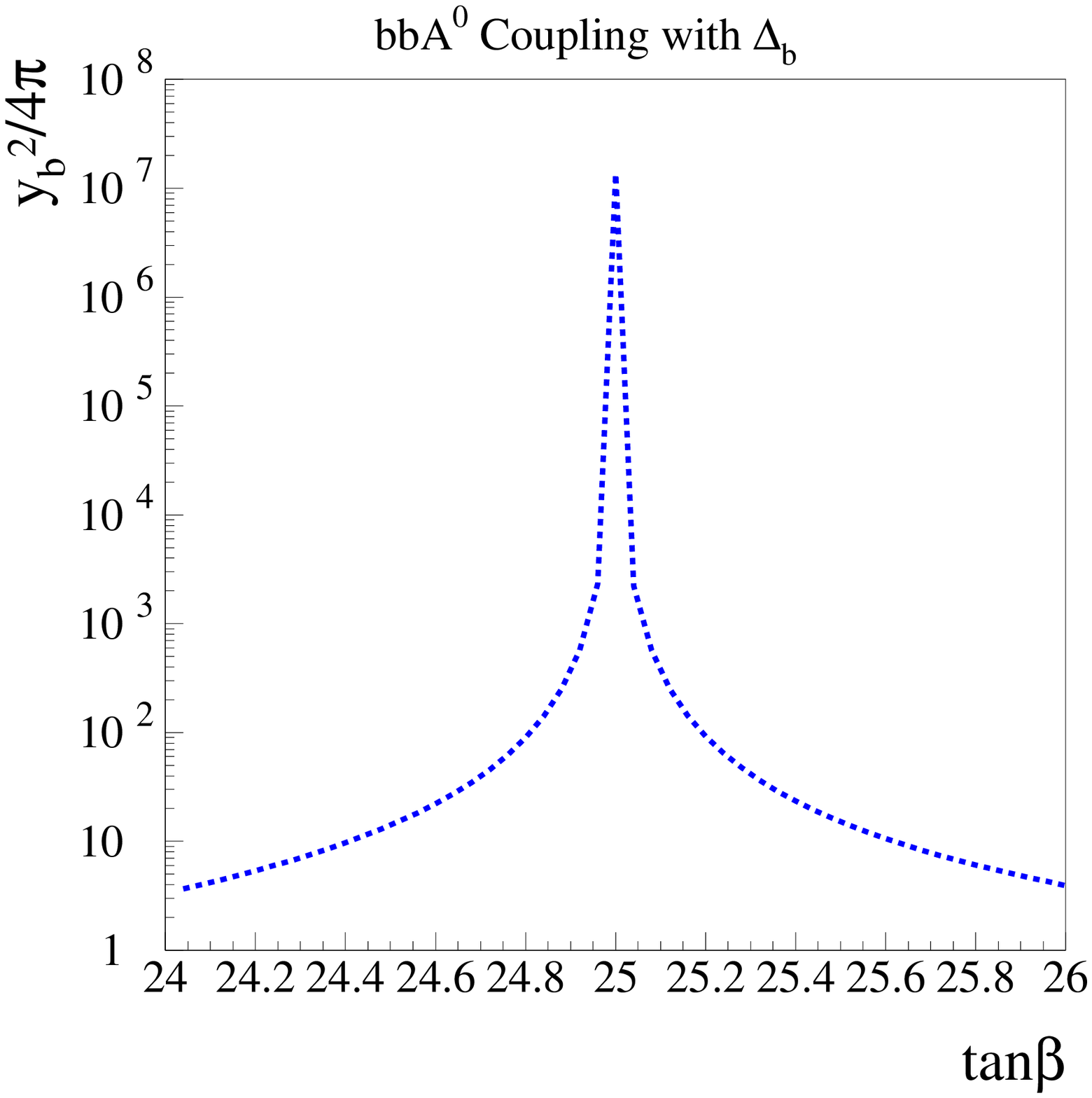}}
\caption{(Left) Dependence of $\sig(\nu\anti\nu\ha)$ on $\tanb$ 
as in Fig.~\ref{deltabe}, for $24 \leq \tan\beta \leq 26$.
(Right) The corresponding value of $y_b^2/(4\pi)$.}
\label{deltabhb}
\end{figure}

Since supersymmetry is broken, the bottom quark
will have, in addition to its usual tree-level coupling to the Higgs 
field $\Phi_1^0$,
a small one-loop-induced coupling to $\Phi_2^0$ 
that couples to up quarks at tree-level:
\begin{equation}
-\mathcal{L}_{\rm Yukawa} \simeq h_b \Phi_1^0 b \anti{b} + (\Delta h_b)
\Phi_2^0 b\anti{b}\,.
\label{couplings}
\end{equation}
When the Higgs doublets acquire their vacuum
expectation values, the bottom quark mass receives
an extra contribution equal to
$(\Delta h_b) v_2$.  Although $\Delta h_b$ is
one-loop suppressed relative to $h_b$,
for sufficiently large values of $\tan\beta$ ($v_2 \gg v_1$)
the contribution to the bottom quark mass of both terms in
Eq.~(\ref{couplings}) may be comparable in size. This induces a
large modification in the tree--level relation,
\begin{equation}
m_b = {h_b v_1\over\sqrt{2}} (1+\Delta_b)\,, \qquad
\label{yukbmass}
\end{equation}
where $\Delta_b \equiv (\Delta h_b)\tan\beta/h_b$.
The function $\Delta_b$ contains two main
contributions: one from a bottom squark--gluino loop
(which depends on the two bottom squark masses $m_{\tilde b_1}$
and $m_{\tilde b_2}$ and the gluino mass $m_{\tilde g}$) and another one
from a
top squark--higgsino loop (which depends on the two top squark masses
$m_{\tilde t_1}$ and $m_{\tilde t_2}$ and the higgsino mass parameter
$\mu$).  The explicit form of $\Delta_b$ at one-loop in the limit of
$M_S \gg m_b$ is given by \cite{deltamb0,deltamb1,deltamb2}:
\begin{equation}
\Delta_b \simeq {2\alpha_s \over 3\pi}
m_{\tilde g}\mu\tan\beta~I(m_{\tilde b_1},
m_{\tilde b_2},m_{\tilde g})
 + {Y_t \over 4\pi} A_t\mu\tan\beta~I(m_{\tilde t_1},m_{\tilde t_2},\mu),
\label{deltamb}
\end{equation}
where $\alpha_s=g_s^2/4\pi$,
$Y_t\equiv h_t^2/4\pi$, and contributions proportional to the
electroweak gauge couplings have been neglected.  
The function $I$ is manifestly positive.
Since the Higgs coupling proportional to $\Delta h_b$ is a
manifestation of the broken supersymmetry in the low energy theory,
$\Delta_b$ does not decouple
in the limit of large supersymmetry breaking masses. Indeed,
if all supersymmetry breaking mass parameters (and $\mu$)
are scaled by a common factor, the correction
$\Delta_b$ remains constant. For our purposes, the important
implication is the modified form of the $b\anti b\ha$ coupling $y_b$
[compare Eq.~(\ref{typeii})]:
\begin{equation}
        y_b = \frac{m_b}{v} \frac{\tan\beta}{1 + \Delta_b},
\end{equation}
where $\Delta_b\propto\tan\beta$ [see Eq.~(\ref{deltamb})].
For appropriate parameter choices with $\mu<0$
(assuming the standard convention of $m_{\tilde g}>0$), $\Delta_b\sim -1$
will occur at some value of moderate to large $\tanb$. At and near this point,
the triangle diagrams in our calculation 
involving the $b \anti b \ha$ coupling [diagram (2) of
Fig.~\ref{fig:WWA} and the $b$-loop cases of diagrams (1) and (2)
of Fig.~\ref{fig:ZZA}]
will be greatly enhanced leading to a very large cross section.
We illustrate this for the specific choices
of $\mha=300\gev$, $\rts=800\gev$,
$\mu=-2\tev$, $m_{\wtil b_R}=525\gev$, 
$\mgl=M_2=m_{\wt \ell_{L,R}}=m_{\wt q_{L,R}}=\msbotl=m_{\wt t_{L,R}}=1\tev$, 
and $A_t=A_b=A_\tau=\mu/\tanb+\sqrt6 \mstopl$ (corresponding to 
maximal-mixing in the stop sector).
Since $A_t\neq 0$ in general, both the gluino--bottom squark and 
higgsino--top squark loops contribute to $\Delta_b$. 
Note that we have chosen sufficiently large masses for
the SUSY particles that the one-loop contributions
to $\epem\to \nu\anti\nu \ha$ involving them will be
very suppressed. Keeping only the $WW$ fusion $t$ and $b$ loop
diagrams [diagrams (1)-(2) of Fig.~\ref{fig:WWA}]
for simplicity, we plot $\sig(\nu\anti\nu \ha)$ as a function of $\tanb$ for
these parameter choices in Fig.~\ref{deltabe}, and compare
to the result that would be obtained without including $\Delta_b$.
A close-up of the $\tan\beta$ region in which $\Delta_b \sim -1$ is shown
in Fig.~\ref{deltabhb}.
The plots show that for $\tanb$ within a few per mil
of the point where $\Delta_b\sim -1$,
the cross section can approach the femtobarn level.
However, this corresponds to an extremely nonperturbative
$b \anti b \ha$ coupling $y_b^2/(4\pi) \sim 10^5$.
Requiring perturbativity, $y_b^2/(4\pi) \lsim \mathcal{O}(1)$,
yields a cross section of order $10^{-4}$ fb, an enhancement
of 1--2 orders of magnitude compared to the cross section without
the $\Delta_b$ effects.

While it would be rather serendipitous for the MSSM parameters
to be within the rather narrow range of $\tanb$
for which $\Delta_b$ is sufficiently near $-1$ to
yield a significantly enhanced $\epem\to\nu\anti\nu\ha$ 
cross section, one cannot simply rule the possibility out.  
Of course, if the one-loop
enhancement is very large, higher loop corrections would
need to be computed to more precisely evaluate the 
magnitude of the cross section enhancement.

\section{\label{sec:conclusions}Conclusions}

Detection of the CP-odd $\ha$ of a CP-conserving two-Higgs-doublet model
via tree-level production mechanisms might not be possible 
due to: (a) the absence of $VV\ha$ ($V=W,Z$) tree-level couplings;
  (b) kinematic
limitations such as $\sqrt{s_{\epem}}<2\mha,\mha+\mhh$;
and/or (c) the small size of the ``Yukawa radiation'' processes yielding
$t\anti t\ha$ and $b\anti b\ha$ final states [as typical
for intermediate $\tanb$ values
in the ``wedge'' region of $(\mha,\tanb)$ parameter space].
These difficulties become magnified in 
models with more than one CP-odd Higgs boson
(the minimal such Higgs sector is that containing
two-doublets plus one-singlet).  Thus, it is generically important
to compute single-$\ha$ production rates deriving from
one-loop diagrams (i.e. not associated with $\ha$ radiation
from a $b$ or $t$ quark) and ascertain the circumstances
under which such processes might allow detection of the $\ha$. In this paper,
we have computed the full (one-loop) cross section 
for $e^+e^- \to \nu \anti \nu \ha$ in the general CP-conserving 2HDM.
Our results are presented in such a way that they can be easily extended
to more complicated models containing a CP-odd Higgs boson.  Complete
formulae for the matrix elements are given in the Appendices.

We have presented numerical results for three cases
in the context of the CP-conserving type-II 2HDM.
The first case considered is that where the Higgs sector
parameters are chosen using the tree-level MSSM Higgs sector constraints.
For this choice, the 2HDM rapidly enters the ``decoupling'' regime
once $\mha>\mz$. For $\tanb>1$, the $\epem\to\nu\anti\nu\ha$ cross
section is typically rather small, especially for $\mha>2\mt$. However,
for $\tanb<1$ we find that $\epem\to\nu\anti\nu\ha$ production
could provide a viable $\ha$ signal, thus covering this important
part of the $(\mha,\tanb)$ ``wedge''
parameter space region in the 2HDM where $\ha$ discovery using
tree-level processes is not possible. In addition, if detected
the $\nu\anti\nu\ha$ rate would provide a very sensitive constraint on
$\tanb$.

For the decoupling 2HDM parameter choices considered above, the
diagrams contributing to $\nu\anti\nu\ha$ production that involve
Higgs self-couplings ($\ha\ha\hl$ and $\ha\ha\hh$) are typically
smaller (often much smaller) in size than $t$- and $b$-loop
diagrams involving $t\anti t\ha$ and $b\anti b\ha$ couplings.
Thus, we have explored alternative 2HDM parameter choices for
which one is far from the decoupling limit and the Higgs 
self-couplings are as large as they can be without violating perturbativity
or precision electroweak constraints. 
We have found that substantial enhancement
of the $\epem\to\nu\anti\nu\ha$ cross section is possible
when $\tanb>10$, possibly sufficient to make the process
marginally observable. While $\ha$ discovery 
via the $\nu\anti\nu\ha$ final state would probably still
be difficult, if the $\ha$ has been detected by other means
and the approximate value of $\tanb$ is already known,
the above results imply that $\epem\to\nu\anti\nu\ha$ might
provide an especially sensitive probe of the Higgs self-couplings
of the 2HDM.

Finally, we considered the effect on the cross section of an 
enhancement of the $b \anti b\ha$ coupling caused by SUSY 
radiative corrections in the MSSM, parameterized by $\Delta_b$.
While the $\Delta_b$ effects can enhance the cross section by 
1--2 orders of magnitude while maintaining a perturbative
$b \anti b \ha$ coupling, this enhancement typically occurs at moderate
to large values of $\tan\beta$ where the cross section is already
quite tiny, so that even with a $\Delta_b$ enhancement the cross section
is not larger than about $10^{-4}$ fb.

For most parameter choices, the $\epem\to\nu\anti\nu\ha$ cross section
is smaller than other single $\ha$ production channels,
most notably $\epem\to \gam\ha$. However, the two
production modes are complementary in important respects.
First, the $\nu\anti\nu\ha$ cross section could provide confirmation
of a signal seen in the $\gam\ha$ final state.
Second, if $\tanb$ is not large, as might
be known either because the two rates are fairly large
or from other Higgs sector measurements, then $\Delta_b$ or
Higgs self-coupling enhancements cannot be substantial and
the determinations of $\tanb$ provided by
the two processes can be fruitfully combined.
Third, while the $\gam\ha$ rate could also be 
enhanced by $\Delta_b$ effects it cannot be enhanced by large
Higgs self-couplings (the needed $\ha\hpm\hmp$  and
$Z\hpm\gmp$ vertices for the $\hp$-loop and $\hp$-$\gp$--loop, respectively, 
being absent in the 2HDM). Thus, an unexpectedly large
cross section in the $\epem\to\nu\anti\nu\ha$ channel would
signal large Higgs self-couplings if a similar enhancement is not
found in the $\gam\ha$ final state. We note that this cross-check
would be important even if no evidence for SUSY particles is found
since a large $\Delta_b$ can arise for arbitrarily large SUSY
particle masses.

An important extension of this work will be to include the
contributions from one-loop diagrams that involve supersymmetric particles.
For a light SUSY spectrum, very substantial enhancements could occur.
This could be especially important in the following situation.
Imagine that the LHC (or Tevatron) discovers fairly light SUSY
particles and a SM-like $\hl$ but is unable to detect the
heavier Higgs bosons $\ha$, $\hh$ and $\hpm$.  
This situation arises at the LHC for moderate $\tan\beta$ values within 
the wedge beginning at $\mha\sim 200\gev$
and becoming rapidly wider in $\tan\beta$ as $\mha$ increases.
If a linear $e^+e^-$ collider has too low a center-of-mass energy
for $\ha\hh$ and $\hp\hm$ pair production, we
must search for the $\ha$ (and the $\hh$ and $\hpm$) 
in the single production modes.
For a light enough SUSY spectrum these modes could be sufficiently
enhanced to make $\gam\ha$, $\nu\anti\nu\ha$ and similar processes 
observable, as found to be the case for $\epem\to \wpm\hmp$ 
production~\cite{Logan:2002jh}.

\begin{acknowledgments}
We thank A.~Arhrib for useful discussions and comparisons of 
numerical results.
T.F. and J.F.G. are supported by U.S. Department of Energy grant
No. DE-FG03-91ER40674 and by the Davis Institute for High
Energy Physics.
H.E.L. is supported in part by the U.S.~Department of Energy
under grant DE-FG02-95ER40896
and in part by the Wisconsin Alumni Research Foundation.
S.S. is supported by the DOE under grant DE-FG03-92-ER-40701 and 
by the John A. McCone Fellowship.
\end{acknowledgments}

\appendix

\section{\label{app:conventions}Notation and conventions}
We follow the notation used on the LoopTools~\cite{Hahn:1998yk}
web page (as of the date of this paper) for the one-loop integrals.
To avoid any possible confusion, our explicit conventions are given below.
The two-point integrals are:
\begin{equation}
        \frac{i}{16\pi^2}\left\{B_0, k^{\mu}B_1 \right\}(k^2,m_1^2,m_2^2)
         =\int \! \frac{d^Dq}{(2\pi)^D} 
        \frac{ \{1, q^{\mu} \} }{(q^2 - m_1^2)((q+k)^2 - m_2^2)},
\end{equation}
where $D$ is the number of dimensions.


\begin{figure}[h]
\setlength{\unitlength}{1.0pt}        
 \begin{center}
\begin{picture}(155,140)(0,13)
\color{blue}
\SetScale{1.0}                        
\ArrowLine(16,25)(40,40)
\ArrowLine(16,135)(40,120)
\ArrowLine(136,80)(105,80)
\ArrowLine(40,40)(40,120)
\ArrowLine(105,80)(40,40)
\ArrowLine(40,120)(105,80)
\Vertex(40,40){2}
\Vertex(105,80){2}
\Vertex(40,120){2}
\Text(14,137)[br]{$p_1$}
\Text(141,80)[cl]{$p_2$}
\Text(14,23)[tr]{$p_3$}

\Text(58,80)[cr]{$m_1$}
\Text(62,90)[bl]{$m_2$}
\Text(62,71)[tl]{$m_3$}
\Text(31,80)[]{$q$}
\Text(85,110)[]{$q+k_1$}
\Text(90,54)[]{$q+k_2$}

\Text(141,29)[]{$k_1=p_1$\hspace{22pt}}
\Text(141,19)[]{$k_2=p_1+p_2$}
\end{picture}
 \end{center}
\caption{Illustration of the LoopTools conventions employed
for the triangle diagram case.}
\label{ltriangle}
\end{figure}
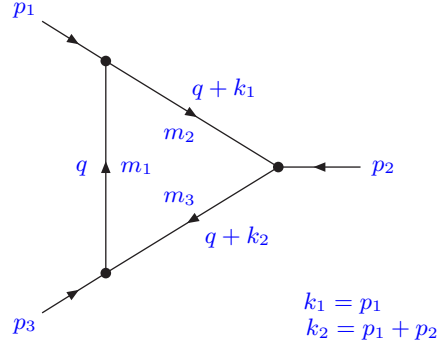


The three-point integrals are:
\begin{equation}
        \frac{i}{16 \pi^2}\left\{ C_0, C^{\mu}, C^{\mu\nu} \right\}
        =
        \hskip0cm \int \! \frac{d^Dq}{(2\pi)^D}
        \frac{ \{1, q^{\mu}, q^{\mu}q^{\nu} \} }
        {(q^2 - m_1^2)((q+k_1)^2 - m_2^2)((q+k_2)^2 - m_3^2)},
        \nonumber
\end{equation}
where the denominator structure follows from the Feynman diagram of
Fig.~\ref{ltriangle}.
The tensor integrals are decomposed in terms of scalar components as
\begin{eqnarray}
        C^{\mu} &=& k_1^{\mu}C_1 + k_2^{\mu}C_2  \nonumber \\
        C^{\mu\nu} &=& g^{\mu\nu} C_{00} + k_1^{\mu}k_1^{\nu}C_{11}
        + k_2^{\mu}k_2^{\nu}C_{22} 
        + (k_1^{\mu}k_2^{\nu} + k_2^{\mu}k_1^{\nu})C_{12}.
\end{eqnarray}
The arguments of the scalar three-point integrals are
specified in our convention
as $(k_1^2,(k_2-k_1)^2,k_2^2,m_1^2,m_2^2,m_3^2)$.

The four-point integrals are:
\begin{eqnarray}
        &&\frac{i}{16 \pi^2}\left\{ D_0, D^{\mu}, D^{\mu\nu} \right\}
        =\hskip0cm \int \! \frac{d^Dq}{(2\pi)^D}
        \frac{ \{1, q^{\mu}, q^{\mu}q^{\nu} \} }
        {(q^2 - m_1^2)((q+k_1)^2 - m_2^2)((q+k_2)^2 - m_3^2)((q+k_3)^2-m_4^2)},
\end{eqnarray}
where the tensor integrals are decomposed in terms of scalar components
as
\begin{eqnarray}
        D^{\mu} &=& k_1^{\mu}D_1 + k_2^{\mu}D_2 + k_3^{\mu}D_3  \nonumber \\
        D^{\mu\nu} &=& g^{\mu\nu} D_{00} + k_1^{\mu}k_1^{\nu}D_{11}
        + k_2^{\mu}k_2^{\nu}D_{22} + k_3^{\mu}k_3^{\nu}D_{33}\nonumber \\
        &&+ (k_1^{\mu}k_2^{\nu} + k_2^{\mu}k_1^{\nu})D_{12}
        + (k_1^{\mu}k_3^{\nu} + k_3^{\mu}k_1^{\nu})D_{13}
        + (k_2^{\mu}k_3^{\nu} + k_3^{\mu}k_2^{\nu})D_{23}.
\end{eqnarray}
The arguments of the scalar four-point integrals are 
$(k_1^2, (k_2-k_1)^2, (k_3-k_2)^2, k_3^2, k_2^2, 
(k_1-k_3)^2, m_1^2, m_2^2, m_3^2, m_4^2)$.

For the three-point functions $C_i$ and $C_{ij}$,
it is useful to define the sums and differences of one-loop integrals
as follows:
\begin{eqnarray}
        C^S_{i,ij}(k_1^2,(k_2-k_1)^2,k_2^2,m_1^2,m_2^2,m_3^2)
        &=& \frac{1}{2} \left[
        C_{i,ij}(k_1^2,(k_2-k_1)^2,k_2^2,m_1^2,m_2^2,m_3^2)
        + C_{i,ij}((k_2-k_1)^2,k_1^2,k_2^2,m_1^2,m_2^2,m_3^2)\right] 
\nonumber \\
        C^D_{i,ij}(k_1^2,(k_2-k_1)^2,k_2^2,m_1^2,m_2^2,m_3^2)
        &=& \frac{1}{2} \left[
        C_{i,ij}(k_1^2,(k_2-k_1)^2,k_2^2,m_1^2,m_2^2,m_3^2)
        - C_{i,ij}((k_2-k_1)^2,k_1^2,k_2^2,m_1^2,m_2^2,m_3^2)\right] 
\nonumber \\
\end{eqnarray}


\section{\label{app:formula_t}2HDM contributions to t-channel diagrams}

We list here our results for the $t$-channel, i.e. $WW$-fusion, diagrams.
In the expressions below, $k_1$ and $k_2$ denote the momenta of the
$\wp$ and $\wm$, respectively, with directions such 
that $k_1=p_\nu-p_{e^-}$ and $k_2=p_{\anti\nu}-p_{e^+}$;
see Fig. \ref{fig:WWA}.1.
(These $k_{1,2}$ should not be confused with those defining
the LoopTools conventions in Appendix \ref{app:conventions}.)

Fig.~\ref{fig:WWA}.1 ($ttb$ loop):
\bea
F_t&=& i{g^3\mt^2\cotb N_c\over 16\mw\pi^2}C_{2}^S,\\
G_t&=&{g^3\mt^2\cotb N_c\over 32\mw\pi^2}\left[(k_1^2-k_2^2)C_{2}^S
-\mha^2 C_{2}^D\right],\\
H_t&=&{g^3\mt^2\cotb N_c\over 16\mw\pi^2}C_{2}^D\,,
\eea
with the arguments for the integral functions as 
$C(k_1^2,k_2^2,\mha^2,\mt^2,\mb^2,\mt^2)$.

Fig.~\ref{fig:WWA}.2 ($bbt$ loop):
\bea
F_b&=& i{g^3\mb^2\tanb N_c\over 16\mw\pi^2}C_{2}^S,\\
G_b&=&-{g^3\mb^2\tanb N_c\over 32\mw\pi^2}\left[(k_1^2-k_2^2)C_{2}^S
-\mha^2 C_{2}^D\right],\\
H_b&=&-{g^3\mb^2\tanb N_c\over 16\mw\pi^2}C_{2}^D\,,
\eea
with the arguments for the integral functions as 
$C(k_1^2,k_2^2,\mha^2,\mb^2,\mt^2,\mb^2)$.

Fig.~\ref{fig:WWA}.3 ($SSS$ loop):
\begin{eqnarray}
        G_{SSS} &=& - \frac{ie^3}{2 \pi^2} g_{AS_1S_3} g_{WS_2S_1} g_{WS_3S_2}
        C^D_{00},
        \\
        H_{SSS} &=& - \frac{ie^3}{2 \pi^2} g_{AS_1S_3} g_{WS_2S_1} g_{WS_3S_2}
        [C^D_{22} + C^D_{12} + C^D_2],
\end{eqnarray}
with the arguments for the integral functions as 
$C(k_1^2,k_2^2,\mha^2,m_1^2,m_2^2,m_3^2)$.

The combinations of scalar particles to be summed over and the
respective couplings are given for the general type-II 2HDM in 
Table~\ref{tab:coups3}.
\begin{table}[h!]
\begin{center}
\begin{tabular}{|ccc|c|c|c|}
\hline
$S_1$ & $S_2$ & $S_3$ & $g_{\ha S_1S_3}$ & $g_{WS_2S_1}$ & $g_{WS_3S_2}$ \\
\hline
$\hl$ & $\hp$ & $\ha$ 
        & $g_{\ha\ha\hl}$
        & $-\cos(\beta - \alpha) / 2 s_W$
        & $-i/2 s_W$ \\
$\hh$ & $\hp$ & $\ha$
        & $g_{\ha\ha\hh}$
        & $\sin(\beta - \alpha) / 2 s_W$
        & $-i/2 s_W$ \\
$\hl$ & $\gp$ & $\go$ 
        & $(\mha^2 - \mhl^2) \cos(\beta - \alpha) / 2 m_W s_W$
        & $-\sin(\beta - \alpha)/2 s_W$
        & $-i/2 s_W$ \\
$\hh$ & $\gp$ & $\go$
        & $(\mhh^2 - \mha^2) \sin(\beta - \alpha) / 2 m_W s_W$
        & $-\cos(\beta - \alpha)/2 s_W$
        & $-i/2 s_W$ \\
$\hm$ & $\hl$ & $\gm$
        & $i(m_{H^{\pm}}^2 - \mha^2)/2 m_W s_W$
        & $\cos(\beta - \alpha)/2 s_W$
        & $\sin(\beta - \alpha)/2 s_W$ \\
$\hm$ & $\hh$ & $\gm$
        & $i(m_{H^{\pm}}^2 - \mha^2)/2 m_W s_W$
        & $-\sin(\beta - \alpha)/2 s_W$
        & $\cos(\beta - \alpha)/2 s_W$ \\
\hline
\end{tabular}
\end{center}
\vspace*{-.2in}
\caption{\label{tab:coups3} The $\ha SS$ and $VSS$ couplings needed for
Fig.~\ref{fig:WWA}.3 are tabulated. Here, $s_W$ and $c_W$ are
the sine and cosine of the Weinberg angle.}
\end{table}
Note that the ``flipped'' diagrams with
$S_1 \leftrightarrow S_3$ have already been taken into account 
in the form factors given above.

In the MSSM, the $\ha\ha\hl$ and $\ha\ha\hh$ coupling coefficients are:
\begin{eqnarray}
        g_{\ha\ha\hl} &=& (-m_Z/2 s_W c_W) \cos 2\beta \sin(\beta + \alpha), 
        \\
        g_{\ha\ha\hh} &=& (m_Z/2 s_W c_W) \cos 2 \beta \cos(\beta + \alpha).
        \label{eq:AAhHMSSM}
\end{eqnarray}
In the general 2HDM, these two coefficients are best expressed \cite{Gunion:2002zf} in terms of certain combinations of the $\lam_i$ 
of Eq.~(\ref{pot}). For $\lam_6=\lam_7=0$, as assumed in this paper,
we have 
\bea
g\ls{\ha\ha\hl} &=&
   -{v\over e}\bigl[\lamT\sbma-\lamU\cbma\bigr]\,, \\[5pt]
g\ls{\ha\ha\hh} &=&  
   -{v\over e}\bigl[\lamT\cbma+\lamU\sbma\bigr]\,,
\label{gs2hdm}
\eea
where ${v\over e}={2c_W\mz\over s_W g^2}$ and
\bea
\lamT &=&
\quarter\stwob^2(\lam_1+\lam_2)+\lamtil(\sb^4+\cb^4)-2\lam_5\,, \label{lamtdef}\\[5pt]
\lamU &=&\half s_{2\beta}(\sb^2\lam_1-\cb^2\lam_2+
c_{2\beta}\lamtil)\,.
\label{lamudef}
\eea
In Eqs.~(\ref{lamtdef}) and (\ref{lamudef}) the
quartic Higgs couplings $\lambda_{1,2,3,4}$ are given in 
terms of the parameter set $\mha$, $\mhh$, $\mhl$, $\mhpm$, $\alpha$, $\tanb$ and $\lam_5$ by Eqs.~(\ref{inverseA})--(\ref{inverseD}) of Sec.~\ref{sec:2HDM}.

Fig.~\ref{fig:WWA}.4$+$Fig.~\ref{fig:WWA}.5  ($SSV$ loop):
\begin{eqnarray}
        G_{SSV} &=& \frac{ie^3}{4 \pi^2} g_{AS_1V_3} g_{WS_2S_1} g_{WV_3S_2}
        C^D_{00},\\
        H_{SSV} &=& \frac{ie^3}{4 \pi^2} g_{AS_1V_3} g_{WS_2S_1} g_{WV_3S_2}
        [C^D_{22} + C^D_{12} - C^D_2],
\end{eqnarray}
with the arguments for the integral functions as 
$C(k_1^2,k_2^2,\mha^2,m_1^2,m_2^2,m_3^2)$.
The combinations of scalar and vector particles to be summed over and the
respective couplings are given in Table~\ref{Table2}.
\begin{table}[h!]
\begin{center}
\begin{tabular}{|ccc|c|c|c|}
\hline
$S_1$ & $S_2$ & $V_3$ & $g_{\ha S_1V_3}$ & $g_{WS_2S_1}$ & $g_{WV_3S_2}$ \\
\hline
$\hm$ & $\hl$ & $W^-$ 
        & $i/2 s_W$
        & $\cos(\beta - \alpha)/2 s_W$
        & $(m_W/s_W)\sin(\beta - \alpha)$ \\
$\hm$ & $\hh$ & $W^-$
        & $i/2 s_W$
        & $-\sin(\beta - \alpha)/2 s_W$
        & $(m_W/s_W)\cos(\beta - \alpha)$ \\
$\hl$ & $\gp$ & $Z$ 
        & $(-i/2s_Wc_W) \cos(\beta - \alpha)$
        & $-\sin(\beta - \alpha)/2 s_W$
        & $-m_Z s_W$ \\
$\hh$ & $\gp$ & $Z$
        & $(i/2s_Wc_W) \sin(\beta - \alpha)$
        & $-\cos(\beta - \alpha)/2 s_W$
        & $-m_Z s_W$ \\
\hline
\end{tabular}
\end{center}
\vspace*{-.2in}
\caption{\label{tab:coups45} The $VSS$ couplings needed for Figs.~\ref{fig:WWA}.4 and \ref{fig:WWA}.5 are tabulated.}
\label{Table2}
\end{table}
Again we have already taken into account the ``flipped'' diagrams with
$S_1 \leftrightarrow V_3$.

Fig.~\ref{fig:WWA}.6$+$Fig.~\ref{fig:WWA}.7  ($SVV$ loop):
\begin{eqnarray}
        G_{SVV} &=& \frac{ie^3}{8 \pi^2} \frac{c_W}{s_W}
        g_{AS_1V_3} g_{WV_2S_1}
        \left\{ [C^D_{00} + 2 k_1 \cdot k_2 C^D_1 
        + \mha^2 C^D_2 - m_1^2 C^D_0 \frac{}{} \right.\nonumber \\
        &&\left. + (k_1^2 - k_2^2) (C^S_0 - C^S_2)]
        + \frac{1}{2}B_0(k_1^2,m_2^2,m_3^2) 
        - \frac{1}{2}B_0(k_2^2,m_2^2,m_3^2) \right\}
        ,\\
        H_{SVV} &=& \frac{ie^3}{8 \pi^2} \frac{c_W}{s_W}
        g_{AS_1V_3} g_{WV_2S_1} 
        [C^D_{22} + C^D_{12} - 4 C^D_1 - C^D_2].
\end{eqnarray}
with the arguments for the integral functions as 
$C(k_1^2,k_2^2,\mha^2,m_1^2,m_2^2,m_3^2)$.
The combinations of scalar and vector particles to be summed over and the
respective couplings are given in Table~\ref{Table3}.
\begin{table}[h!]
\begin{center}
\begin{tabular}{|ccc|c|c|}
\hline
$S_1$ & $V_2$ & $V_3$ & $g_{\ha S_1V_3}$ & $g_{WV_2S_1}$ \\
\hline
$\hl$ & $W^+$ & $Z$
        & $(-i/2 s_W c_W) \cos(\beta - \alpha)$
        & $(m_W/s_W) \sin(\beta - \alpha)$ \\
$\hh$ & $W^+$ & $Z$
        & $(i/2 s_W c_W) \sin(\beta - \alpha)$
        & $(m_W/s_W) \cos(\beta - \alpha)$ \\
\hline
\end{tabular}
\end{center}
\vspace*{-.2in}
\caption{\label{tab:coups67} The $\ha S V$ and $WVS$ couplings
needed for Figs.~\ref{fig:WWA}.6 and \ref{fig:WWA}.7 are tabulated.}
\label{Table3}
\end{table}
Again we have already taken into account the ``flipped'' diagrams with
$S_1 \leftrightarrow V_3$.

It is easy to check that in the on-shell limit where 
$k_1^2=k_2^2=m_W^2$, all the $C_{i,ij}^D$'s go to zero.  Contributions 
to $G$ and $H$ vanish and the only contribution to $\wp\wm\ha$ effective 
coupling comes from $F$, as pointed out in \cite{gunhabkao}.

Fig.~\ref{fig:WWA}.8 (box diagrams):
\bea
G&=&\frac{g^3}{16 \pi^2}g_Z^\nu g_{WWS} g_{ZSA} (k_2^2-m_W^2)
\left[
C_0+D_0(m_Z^2+2\sp+2t_2+2u_1)\right.
\nonumber \\
&&\left.+D_1(2\mha^2+\sp+t_2+u_1)
+D_2(2\mha^2-2s-2t_1-2u_2+t_2)+D_3(2\sp+2t_2+2u_1)\right]
\\
H_1=H_4&=&\frac{g^3}{16 \pi^2}g_Z^\nu g_{WWS} g_{ZSA} (k_2^2-m_W^2)(4D_2),
\eea
with the arguments for the integral functions as 
$D(\mha^2,t_1,0,0,t_2,s+t_1+u_2,m_Z^2,m_S^2,m_W^2,0)$,
$C_0(t_1,0,s+t_1+u_2,m_S^2,m_W^2,0)$, $g_Z^\nu=-T_{3\nu}/c_W$,
and
\beq
s=2p_{e^-}\cdot p_{\ep},\quad\sp=2p_{\nu}\cdot p_{\anti\nu},\quad t_1=k_1^2=-2p_{e^-}\cdot p_{\nu},\quad t_2=k_2^2=-2p_{\ep}\cdot p_{\anti\nu},\quad
u_1=-2p_{e^-}\cdot p_{\anti\nu},\quad u_2=-2p_{\ep}\cdot p_{\nu}\,.
\eeq
The scalar could be $\hl$ or $\hh$, with the couplings given in
Table~\ref{Table4}.
\begin{table}[h!]
\begin{center}
\begin{tabular}{|c|c|c|}
\hline
$S$ & $g_{WWS}$ & $g_{ZS\ha}$ \\
\hline
$\hl$ & $m_W\sin(\beta-\alpha)$ & $\cos(\beta-\alpha)/(2c_W)$\\
$\hh$ & $m_W\cos(\beta-\alpha)$ & $-\sin(\beta-\alpha)/(2c_W)$\\
\hline
\end{tabular}
\end{center}
\vspace*{-.2in}
\caption{\label{tab:coups89} The $WWS$ and $ZS\ha$ couplings needed
for Figs.~\ref{fig:WWA}.8--\ref{fig:WWA}.11 are given.}
\label{Table4}
\end{table}

Fig.~\ref{fig:WWA}.9: Similar to Fig.\ref{fig:WWA}.8, under the exchange of 
\bea
&&H_4\rightarrow H_3,\ \ \ u_1\leftrightarrow u_2,\ \ \  
t_1\leftrightarrow t_2,\ \ \ 
k_1^2\leftrightarrow k_2^2,
\ \ \ {\rm and\  an\  overall\  ``-"\  sign}.
\eea

Fig.~\ref{fig:WWA}.10: Similar to Fig.\ref{fig:WWA}.8, under the exchange of 
\bea
&&g_Z^\nu\rightarrow g_Z^{eL}=-(T_{3eL}-Q_es_W^2)/c_W,\  
H_1\rightarrow H_2,\ H_4\rightarrow H_3,\  
s\leftrightarrow\sp,\  u_1\leftrightarrow u_2,
\ {\rm and\  an\  overall\  ``-"\ sign}.
\eea

Fig.~\ref{fig:WWA}.11: Similar to Fig.~\ref{fig:WWA}.8, under the exchange of 
\bea
&&g_Z^\nu\rightarrow g_Z^{eL}=-(T_{3eL}-Q_es_W^2)/c_W,\ \ \ 
H_1\rightarrow H_2,\ \ \
s\leftrightarrow\sp,\ \ \  t_1\leftrightarrow t_2,\ \ \ 
k_1^2\leftrightarrow k_2^2.
\eea

\section{\label{app:formula_s}2HDM contributions to s-channel diagrams}

For the $s$-channel diagrams, we introduce the following operators:
\begin{equation}
        \begin{array}{lcl}
\mathcal{O}_1 = \anti v(p_{e^+}) \gamma^{\mu} P_R u(p_{e^-})
        \anti u(p_{\nu}) \gamma_{\mu} P_L v(p_{\anti \nu}) 
&      \quad &
\mathcal{O}_2 = \anti v(p_{e^+}) \gamma^{\mu} P_L u(p_{e^-})
        \anti u(p_{\nu}) \gamma_{\mu} P_L v(p_{\anti \nu})  \\
\mathcal{O}_3 = \anti v(p_{e^+}) \not{p_{\nu}} P_R u(p_{e^-})
        \anti u(p_{\nu}) \not{p_{e^-}} P_L v(p_{\anti \nu}) 
&      \quad &
\mathcal{O}_4 = \anti v(p_{e^+}) \not{p_{\nu}} P_L u(p_{e^-})
        \anti u(p_{\nu}) \not{p_{e^-}} P_L v(p_{\anti \nu})  \\
\mathcal{O}_5 = \anti v(p_{e^+}) \not{p_{\nu}} P_R u(p_{e^-})
        \anti u(p_{\nu}) \not{p_{e^+}} P_L v(p_{\anti \nu}) 
&      \quad &
\mathcal{O}_6 = \anti v(p_{e^+}) \not{p_{\nu}} P_L u(p_{e^-})
        \anti u(p_{\nu}) \not{p_{e^+}} P_L v(p_{\anti \nu})  \\
\mathcal{O}_7 = \anti v(p_{e^+}) \not{p_{\anti \nu}} P_R u(p_{e^-})
        \anti u(p_{\nu}) \not{p_{e^-}} P_L v(p_{\anti \nu}) 
&      \quad &
\mathcal{O}_8 = \anti v(p_{e^+}) \not{p_{\anti \nu}} P_L u(p_{e^-})
        \anti u(p_{\nu}) \not{p_{e^-}} P_L v(p_{\anti \nu})  \\
\mathcal{O}_9 = \anti v(p_{e^+}) \not{p_{\anti \nu}} P_R u(p_{e^-})
        \anti u(p_{\nu}) \not{p_{e^+}} P_L v(p_{\anti \nu}) 
&      \quad &
\mathcal{O}_{10} = \anti v(p_{e^+}) \not{p_{\anti \nu}} P_L u(p_{e^-})
        \anti u(p_{\nu}) \not{p_{e^+}} P_L v(p_{\anti \nu}).
        \end{array}
\end{equation}

We now list our results for $s$-channel diagrams.

Fig.~\ref{fig:ZZA}.1+\ref{fig:ZZA}.2 ($s$-channel top quark loop): 
\begin{eqnarray}
        \mathcal{M}&=& 
        \frac{2 \alpha^2 N_c g_Z^{\nu L}}
        {(s - m_V^2)(\hat s - m_Z^2 + i m_Z \Gamma_Z)}
        \frac{e m_t^2 \cot\beta}{m_W s_W} 
        \left[ (g_V^{tR} g_Z^{tL} + g_V^{tL} g_Z^{tR}) C_2
        - (g_V^{tL} g_Z^{tL} + g_V^{tR} g_Z^{tR})(C_0 + C_2) \right]
        \nonumber \\
        &&\!\!\!\!\!
        \times \left[ 
        \frac{1}{2} (t_1 - t_2 + u_1 - u_2)
        (-g_V^{eR} \mathcal{O}_1 + g_V^{eL} \mathcal{O}_2) 
        - g_V^{eR} (\mathcal{O}_3 + \mathcal{O}_7 
        - \mathcal{O}_5 - \mathcal{O}_9)
        + g_V^{eL} (\mathcal{O}_4 + \mathcal{O}_8
        - \mathcal{O}_6 - \mathcal{O}_{10}) \right],
\end{eqnarray}
with the arguments for the integral functions as 
$C(\mha^2, \hat s, s, m_t^2, m_t^2, m_t^2)$.
This includes the top quark going clockwise and counterclockwise.
The gauge boson connecting the initial $e^+e^-$ to the top quark loop
is $V = \gamma$ or $Z$. 
The couplings are defined as:
\begin{eqnarray}
        g_Z^{qL} &=& \frac{(-T_3^q + Q_q s^2_W)}{s_Wc_W}, \qquad
        g_Z^{qR} = \frac{Q_q s^2_W}{s_Wc_W}, \qquad
        g_{\gamma}^{qL} = g_{\gamma}^{qR} = -Q_q.
\end{eqnarray}
To get the $s$-channel bottom quark loop, make the following substitutions:
\begin{equation}
        m_t \to m_b, \qquad \cot\beta \to \tan\beta,
        \qquad (g_V^{tL}, g_V^{tR}) \to (g_V^{bL}, g_V^{bR}).
\end{equation}

Fig.~\ref{fig:ZZA}.3 (box diagram): For $\hl$ in the box,
\begin{eqnarray}
        \mathcal{M} &=& 
        \frac{\alpha^2 \sin 2(\beta - \alpha) m_W}{4 s^2_W c^3_W}
        \frac{e g^{\nu L}_Z}
        {(\hat s - m_Z^2 + i m_Z \Gamma_Z)} \times \left[
        ((g_Z^{eR})^2 \mathcal{O}_1 + (g_Z^{eL})^2 \mathcal{O}_2)
        \right.(-C_0 + (\hat s + t_1 + u_1 - \mhl^2) D_0 
        \nonumber \\
        &&\left. 
        + (t_1 + u_1) D_1
        + (\mha^2 - \hat s - t_1 - u_1) D_3) 
        + 4 D_1 ((g_Z^{eR})^2 (\mathcal{O}_3 + \mathcal{O}_7)
                + (g_Z^{eL})^2 (\mathcal{O}_4 + \mathcal{O}_8))
        \right],
\end{eqnarray}
with the arguments for the integral functions as 
$C(0,0,s,m_Z^2,0,m_Z^2)$ and 
$D(\hat s,0,0,\mha^2,(\hat s + t_1 + u_1),s,\mhl^2,m_Z^2,0,m_Z^2)$.

The crossed box is obtained by applying the substitutions:
\begin{equation}
        t_1 \to t_2, \quad u_1 \to u_2, \quad
        (\mathcal{O}_3, \mathcal{O}_4, \mathcal{O}_7, \mathcal{O}_8)
        \to (\mathcal{O}_5, \mathcal{O}_6, \mathcal{O}_9, \mathcal{O}_{10}),
        \quad 
        \mbox{\rm and~an~overall~``$-$''~sign}.
\end{equation}
The box diagram containing $\hh$ is obtained by applying the substitution:
\begin{equation}
        \mhl \to \mhh \qquad \mbox{\rm and~an~overall~``$-$''~sign}.
\end{equation}

Fig.~\ref{fig:ZZA}.4: For $\hl$ in the box,
\begin{eqnarray}
        \mathcal{M} &=& 
        \frac{\alpha^2 \sin 2(\beta - \alpha) m_W}{4 s^2_W c^3_W}
        \frac{e (g^{\nu L}_Z)^2} {(s - m_Z^2)} \left[
        (g_Z^{eR} \mathcal{O}_1 + g_Z^{eL} \mathcal{O}_2)
        (-C_0 + (s + t_2 + u_1 - \mhl^2) D_0 \right. 
        \nonumber \\
        &&\left.
        + (t_2 + u_1) D_1 + (\mha^2 - s - t_2 - u_1) D_3) 
        + 4 D_1 (g_Z^{eR}(\mathcal{O}_7 + \mathcal{O}_9)
        + g_Z^{eL} (\mathcal{O}_8 + \mathcal{O}_{10})
        \right],
\end{eqnarray}
with the arguments for the integral functions as 
$C(0,0,\hat s,m_Z^2,0,m_Z^2)$ and 
$D(s,0,0,\mha^2,u^{\prime},\hat s,\mhl^2,m_Z^2,0,m_Z^2)$, 
where $u^{\prime} = s + t_2 + u_1$.

The crossed box is obtained by applying the substitutions:
\begin{equation}
        t_2 \to t_1, \quad u_1 \to u_2, \quad
        (\mathcal{O}_7, \mathcal{O}_8, \mathcal{O}_9, \mathcal{O}_{10}) \to
        (\mathcal{O}_3, \mathcal{O}_4, \mathcal{O}_5, \mathcal{O}_6),
        \quad \mbox{\rm and~an~overall~``$-$''~sign}.
\end{equation}
The box diagram containing $\hh$ is obtained by applying the substitution:
\begin{equation}
        \mhl \to \mhh \qquad \mbox{\rm and~an~overall~``$-$''~sign}.
\end{equation}

\section{\label{app:square}Square of the matrix element}
The cross section for $e^+e^-\rightarrow \nu \anti{\nu}\ha$ is evaluated 
by integrating the spin-averaged matrix element square 
[Eq.~(\ref{eq:maxelesqu_new})] over the three body phase space of the 
final states:
\bea
{1\over 4} {\sum}_{spins}|\calm|^2&=& {1\over 4} {\sum}_{spins}|\calm_t|^2
+{3\over 4} {\sum}_{spins}|\mathcal{M}_s(e^-_L e^+_R)|^2
+{3\over 4} {\sum}_{spins}|\mathcal{M}_s(e^-_R e^+_L)|^2\nonumber \\
&+&{1\over 4} {\sum}_{spins}\calm_t\mathcal{M}_s(e^-_R e^+_L)^*
+{\rm h.c.},
\label{eq:maxelesqu_new}
\eea
where the 3 in the second and third terms represents the sum over 
three neutrino flavors for the $s$-channel contribution.
The various pieces in Eq.~(\ref{eq:maxelesqu_new}) are given
explicitly below. 

The spin-summed amplitude squared for $t$-channel diagrams is
\def\sp{\what s}
\def\em{e^-}
\def\ep{e^+}
\def\nn{\nu}
\def\nc{\overline\nu}
\bea
{\sum}_{spins}|\calm_t|^2&=& 
{1\over 4}\left\vert{ g^2 \over (t_1-\mw^2)(t_2-\mw^2)}\right\vert^2 K_t
\eea
where
\bea
K_t&=& |F|^2
(-s^2 \sp^2+
     2 s \sp t_1 t_2+2 s \sp u_1 u_2-t_1^2 t_2^2+t_1 t_2 u_1^2+t_1 t_2 u_2^2
-u_1^2 u_2^2)
\nonumber\\
&+&4|G|^2 u_1u_2
\nonumber\\
&+&|H_1|^2s^2 u_1 u_2+|H_2|^2\sp^2 u_1 u_2+|H_3|^2s \sp u_1^2+
|H_4|^2s \sp u_2^2 
\nonumber \\
&+&2\Re(H_1H_2^*)s\sp u_1 u_2
\nonumber \\
&+&\Re(H_1H_3^*)s u_1(s\sp+u_1 u_2-t_1 t_2)
-4\Im(H_1H_3^*)\eps(\em,\ep,\nc,\nn)s u_1
\nonumber \\
&+&\Re(H_1H_4^*)s u_2(s\sp+u_1 u_2-t_1 t_2)
+4\Im(H_1H_4^*)\eps(\em,\ep,\nc,\nn)s u_2
\nonumber \\
&+&\Re(H_2H_3^*)\sp u_1(s\sp+u_1 u_2-t_1 t_2)
-4\Im(H_2H_3^*)\eps(\em,\ep,\nc,\nn)\sp u_1
\nonumber \\
&+&\Re(H_2H_4^*)\sp u_2(s\sp+u_1 u_2-t_1 t_2)
+4\Im(H_2H_4^*)\eps(\em,\ep,\nc,\nn)\sp u_2
\nonumber \\
&+&\Re(H_3H_4^*)(s^2\sp^2-2 s\sp t_1 t_2+t_1^2 t_2^2-2 t_1t_2u_1u_2+u_1^2u_2^2)
+4\Im(H_3H_4^*)\eps(\em,\ep,\nc,\nn)(s\sp+u_1 u_2-t_1 t_2)
\nonumber \\
&+&\Re(FG^*) 8 \eps(\em,\ep,\nc,\nn) (u_1+u_2)
\nonumber\\
&+&2\Im(FG^*)(-s \sp u_1+s\sp u_2+t_1 t_2 u_1-t_1 t_2 u_2+u_1^2 u_2-u_1 u_2^2)
\nonumber\\
&+&4 \Re(FH_1^*)\eps(\em,\ep,\nc,\nn)(s u_1+s u_2)
+\Im(FH_1^*)(-s^2 \sp u_1+s^2 \sp u_2+s u_1^2 u_2-s u_1 u_2^2
+s t_1 t_2 u_1-s t_1 t_2 u_2)
\nonumber \\
&+&4 \Re(FH_2^*)\eps(\em,\ep,\nc,\nn)(\sp u_1+\sp u_2)
+\Im(FH_2^*)(-s \sp^2 u_1+s \sp^2 u_2+\sp u_1^2 u_2-\sp u_1 u_2^2
+\sp t_1 t_2 u_1-\sp t_1 t_2 u_2)
\nonumber \\
&+&4 \Re(FH_3^*)\eps(\em,\ep,\nc,\nn)(s \sp+u_1^2-t_1 t_2)
\nonumber \\
&+&\Im(FH_3^*)(u_1^3 u_2-t_1 t_2 u_1^2-s \sp u_1^2-t_1 t_2 u_1 u_2 
+t_1^2 t_2^2 + s^2 \sp^2 -2 s \sp t_1 t_2- s \sp u_1 u_2)
\nonumber \\
&+&4 \Re(FH_4^*)\eps(\em,\ep,\nc,\nn)(s \sp+u_2^2-t_1 t_2)
\nonumber \\
&-&\Im(FH_4^*)(u_1 u_2^3-t_1 t_2 u_2^2-s \sp u_2^2-t_1 t_2 u_1 u_2 
+t_1^2 t_2^2 + s^2 \sp^2 -2 s \sp t_1 t_2- s \sp u_1 u_2)
\nonumber \\
&+&4 \Re(GH_1^*)s u_1 u_2+4 \Re(GH_2^*)\sp u_1 u_2
\nonumber \\
&+&2\Re(GH_3^*)u_1(u_1 u_2-t_1 t_2 +s \sp)
-8\Im(GH_3^*)\eps(\em,\ep,\nc,\nn)u_1
\nonumber \\
&+&2\Re(GH_4^*)u_2(u_1 u_2-t_1 t_2 +s \sp )
+8\Im(GH_4^*)\eps(\em,\ep,\nc,\nn)u_2,
\eea
and
\beq
\eps(\em,\ep,\nc,\nn)=\eps^{\mu\lam\rho\sigma}p^{\em}_{\mu}p^{\ep}_{\lam}p^{\anti\nu}_{\rho}p^{\nu}_{\sigma}\,,
\eeq
with $\eps^{0123}=+1$.
Here, $\Re$ and $\Im$ denote the real and imaginary parts, respectively,
of the indicated products.

The spin-summed amplitudes-squared for $s$-channel diagrams are
\begin{eqnarray}
        \sum_{spins} |\mathcal{M}_s(e^-_R e^+_L)|^2
        &=& 4 |\mathcal{M}_1|^2 t_1 t_2 
        + |\mathcal{M}_3|^2 t_1^2 u_1 u_2
        + |\mathcal{M}_5|^2 t_1 t_2 u_2^2 
        + |\mathcal{M}_7|^2 t_1 t_2 u_1^2
        + |\mathcal{M}_9|^2 t_2^2 u_1 u_2 \nonumber \\
        &+& 2 {\Re}(\mathcal{M}_1 \mathcal{M}_3^*)
        t_1(s \hat s - t_1 t_2 - u_1 u_2)
        - 8 {\Im}(\mathcal{M}_1 \mathcal{M}_3^*) 
        t_1 \epsilon(e^-,e^+,\anti \nu,\nu) \nonumber \\
        &-& 4 {\Re}(\mathcal{M}_1 \mathcal{M}_5^*) t_1 t_2 u_2
        - 4 {\Re}(\mathcal{M}_1 \mathcal{M}_7^*) t_1 t_2 u_1 \nonumber \\
        &+& 2 {\Re}(\mathcal{M}_1 \mathcal{M}_9^*)
        t_2(s \hat s - t_1 t_2 - u_1 u_2)
        + 8 {\Im}(\mathcal{M}_1 \mathcal{M}_9^*)
        t_2 \epsilon(e^-,e^+,\anti \nu,\nu) \nonumber \\
        &-& {\Re}(\mathcal{M}_3 \mathcal{M}_5^*)
        t_1 u_2 (s \hat s - t_1 t_2 - u_1 u_2)
        - 4 {\Im}(\mathcal{M}_3 \mathcal{M}_5^*) 
        t_1 u_2 \epsilon(e^-,e^+,\anti \nu,\nu) \nonumber \\
        &-& {\Re}(\mathcal{M}_3 \mathcal{M}_7^*)
        t_1 u_1 (s \hat s - t_1 t_2 - u_1 u_2)
        - 4 {\Im}(\mathcal{M}_3 \mathcal{M}_7^*)
        t_1 u_1 \epsilon(e^-,e^+,\anti \nu,\nu) \nonumber \\
        &+& {\Re}(\mathcal{M}_3 \mathcal{M}_9^*)
        \left[\frac{1}{2} (s \hat s - t_1 t_2 - u_1 u_2)^2 
        - 8 (\epsilon(e^-,e^+,\anti \nu,\nu))^2 \right] \nonumber \\
        &+& 4 {\Im}(\mathcal{M}_3 \mathcal{M}_9^*)
        (s \hat s - t_1 t_2 - u_1 u_2) \epsilon(e^-,e^+,\anti \nu,\nu) 
        \nonumber \\
        &+& {\Re}(\mathcal{M}_5 \mathcal{M}_7^*)
        \left[\frac{1}{2} (s \hat s - t_1 t_2 - u_1 u_2)^2 
        + 8 (\epsilon(e^-,e^+,\anti \nu,\nu))^2 \right] \\
        &-& {\Re}(\mathcal{M}_5 \mathcal{M}_9^*)
        t_2 u_2 (s \hat s - t_1 t_2 - u_1 u_2)
        - 4 {\Im}(\mathcal{M}_5 \mathcal{M}_9^*)
        t_2 u_2 \epsilon(e^-,e^+,\anti \nu,\nu) \nonumber \\
        &-& {\Re}(\mathcal{M}_7 \mathcal{M}_9^*)
        t_2 u_1 (s \hat s - t_1 t_2 - u_1 u_2)
        - 4 {\Im}(\mathcal{M}_7 \mathcal{M}_9^*)
        t_2 u_1 \epsilon(e^-,e^+,\anti \nu,\nu) \nonumber 
\end{eqnarray}
\begin{eqnarray}
        \sum_{spins} |\mathcal{M}_s(e^-_L e^+_R)|^2
        &=& 4 |\mathcal{M}_2|^2 u_1 u_2
        + |\mathcal{M}_4|^2 t_1^2 u_1 u_2
        + |\mathcal{M}_6|^2 t_1 t_2 u_2^2 
        + |\mathcal{M}_8|^2 t_1 t_2 u_1^2
        + |\mathcal{M}_{10}|^2 t_2^2 u_1 u_2 \nonumber \\
        &-& 4 {\Re}(\mathcal{M}_2 \mathcal{M}_4^*) t_1 u_1 u_2
        \nonumber \\
        &+& 2 {\Re}(\mathcal{M}_2 \mathcal{M}_6^*)
        u_2 (s \hat s - t_1 t_2 - u_1 u_2)
        + 8 {\Im}(\mathcal{M}_2 \mathcal{M}_6^*)
        u_2 \epsilon(e^-,e^+,\anti\nu,\nu) \nonumber \\
        &+& 2 {\Re}(\mathcal{M}_2 \mathcal{M}_8^*)
        u_1 (s \hat s - t_1 t_2 - u_1 u_2)
        - 8 {\Im}(\mathcal{M}_2 \mathcal{M}_8^*)
        u_1 \epsilon(e^-,e^+,\anti\nu,\nu) \nonumber \\
        &-& 4 {\Re}(\mathcal{M}_2 \mathcal{M}_{10}^*)
        t_2 u_1 u_2 \nonumber \\
        &-& {\Re}(\mathcal{M}_4 \mathcal{M}_6^*)
        t_1 u_2 (s \hat s - t_1 t_2 - u_1 u_2)
        - 4 {\Im}(\mathcal{M}_4 \mathcal{M}_6^*)
        t_1 u_2 \epsilon(e^-,e^+,\anti\nu,\nu) \nonumber \\
        &-& {\Re}(\mathcal{M}_4 \mathcal{M}_8^*)
        t_1 u_1 (s \hat s - t_1 t_2 - u_1 u_2)
        + 4 {\Im}(\mathcal{M}_4 \mathcal{M}_8^*)
        t_1 u_1 \epsilon(e^-,e^+,\anti\nu,\nu) \nonumber \\
        &+& {\Re}(\mathcal{M}_4 \mathcal{M}_{10}^*)
        \left[ \frac{1}{2} (s \hat s - t_1 t_2 - u_1 u_2)^2
        + 8 (\epsilon(e^-,e^+,\anti\nu,\nu))^2 \right] \nonumber \\
        &+& {\Re}(\mathcal{M}_6 \mathcal{M}_8^*)
        \left[ \frac{1}{2} (s \hat s - t_1 t_2 - u_1 u_2)^2
        - 8 (\epsilon(e^-,e^+,\anti\nu,\nu))^2 \right] \nonumber \\
        &-& 4 {\Im}(\mathcal{M}_6 \mathcal{M}_8^*)
        (s \hat s - t_1 t_2 - u_1 u_2) \epsilon(e^-,e^+,\anti\nu,\nu) \\ 
        &-& {\Re}(\mathcal{M}_6 \mathcal{M}_{10}^*)
        t_2 u_2 (s \hat s - t_1 t_2 - u_1 u_2)
        + 4 {\Im}(\mathcal{M}_6 \mathcal{M}_{10}^*)
        t_2 u_2 \epsilon(e^-,e^+,\anti\nu,\nu) \nonumber \\
        &-& {\Re}(\mathcal{M}_8 \mathcal{M}_{10}^*)
        t_2 u_1 (s \hat s - t_1 t_2 - u_1 u_2)
        - 4 {\Im}(\mathcal{M}_8 \mathcal{M}_{10}^*)
        t_2 u_1 \epsilon(e^-,e^+,\anti\nu,\nu), \nonumber
\end{eqnarray}
where $8(\epsilon(e^-,e^+,\anti\nu,\nu))^2=-\half 
(s \hat s + u_1 u_2 - t_1 t_2)^2+2 s \hat s  u_1 u_2$.

The interference terms between $s$- and $t$-channel diagrams are
\begin{equation}
{\sum}_{spins}\calm_t\mathcal{M}_s(e^-_L e^+_R)^*
+{\rm h.c.}=
        \sum_{i \ {\rm even}} \sum_{spins}\mathcal{M}_t 
        \mathcal{M}_i^* \mathcal{O}_i^*
         + {\rm h.c.}
        = \sum_{i \ {\rm even}}
        \left[ \frac{g^2}{2} \frac{1}{(t_1 - m_W^2)(t_2 - m_W^2)} \right]
        K^{\prime}_i ,
\end{equation}
where
\begin{eqnarray}
K^{\prime}_2 &=&
        - 8 {\Re}(G \mathcal{M}_2^*) u_1 u_2
        - 8 {\Re}(F \mathcal{M}_2^*) (u_1 + u_2) 
                \epsilon(e^-,e^+,\anti \nu,\nu) \nonumber \\
        && - 2 {\Im}(F \mathcal{M}_2^*) (u_1 - u_2) 
                (-s \hat s + t_1 t_2 + u_1 u_2) 
        - 4 {\Re}(H_1 \mathcal{M}_2^*) s u_1 u_2 \nonumber \\
        &&
        - 4 {\Re}(H_2 \mathcal{M}_2^*) \hat s u_1 u_2
        - 2 {\Re}(H_3 \mathcal{M}_2^*) u_1 (u_1 u_2 - t_1 t_2 + s \hat s)
        \nonumber \\  &&
        - 8 {\Im}(H_3 \mathcal{M}_2^*) u_1 \epsilon(e^-,e^+,\anti \nu,\nu)
        - 2 {\Re}(H_4 \mathcal{M}_2^*) u_2 (u_1 u_2 - t_1 t_2 + s \hat s)
        \nonumber \\  &&
        + 8 {\Im}(H_4 \mathcal{M}_2^*) u_2 \epsilon(e^-,e^+,\anti \nu,\nu)
        \nonumber \\
K^{\prime}_4 &=& 4 {\Re}(F \mathcal{M}_4^*)
                t_1 (u_1 + u_2)
                \epsilon(e^-,e^+,\anti\nu,\nu)
        \nonumber \\
        && - {\Im}(F \mathcal{M}_4^*)
                t_1 (s \hat s - t_1 t_2 - u_1 u_2) (u_1 - u_2)
        \nonumber \\
        && + 4 {\Re}(G \mathcal{M}_4^*) t_1 u_1 u_2
        + 2 {\Re}(H_1 \mathcal{M}_4^*) s t_1 u_1 u_2
        + 2 {\Re}(H_2 \mathcal{M}_4^*) \hat s t_1 u_1 u_2
        \nonumber \\
        && + {\Re}(H_3 \mathcal{M}_4^*) t_1 u_1 
                (s \hat s + u_1 u_2 - t_1 t_2)
        + 4 {\Im}(H_3 \mathcal{M}_4^*) t_1 u_1 \epsilon(e^-,e^+,\anti\nu,\nu)
        \nonumber \\
        && + {\Re}(H_4 \mathcal{M}_4^*) t_1 u_2
                (s \hat s + u_1 u_2 - t_1 t_2)
        - 4 {\Im}(H_4 \mathcal{M}_4^*) t_1 u_2 \epsilon(e^-,e^+,\anti\nu,\nu)
        \nonumber \\
K^{\prime}_6 &=& 4 {\Re}(F \mathcal{M}_6^*)(t_1 t_2 + u_1 u_2 - s \hat s)
                \epsilon(e^-,e^+,\anti \nu,\nu)
        \nonumber \\
        && + {\Im}(F \mathcal{M}_6^*)[(t_1 t_2 + u_1 u_2 - s \hat s)^2
                - 2 t_1 t_2 u_2 (u_1 + u_2)]
        \nonumber \\
        && + 2 {\Re}(G \mathcal{M}_6^*) u_2 (t_1 t_2 + u_1 u_2 - s \hat s)
        - 8 {\Im}(G \mathcal{M}_6^*) u_2 \epsilon(e^-,e^+,\anti \nu,\nu)
        \nonumber \\
        && + {\Re}(H_1 \mathcal{M}_6^*) s u_2 (t_1 t_2 + u_1 u_2 - s \hat s)
        - 4 {\Im}(H_1 \mathcal{M}_6^*) s u_2 \epsilon(e^-,e^+,\anti \nu,\nu)
        \nonumber \\
        && + {\Re}(H_2 \mathcal{M}_6^*) \hat s u_2 
                (t_1 t_2 + u_1 u_2 - s \hat s)
        - 4 {\Im}(H_2 \mathcal{M}_6^*) \hat s u_2 
                \epsilon(e^-,e^+,\anti \nu,\nu)
        \nonumber \\
        && + {\Re}(H_3 \mathcal{M}_6^*) 
                [-(t_1 t_2 - s \hat s)^2 + u_1 u_2 (t_1 t_2 + s \hat s)]
        \nonumber \\
        && + 4 {\Im}(H_3 \mathcal{M}_6^*) (t_1 t_2 - s \hat s)
                \epsilon(e^-,e^+,\anti \nu,\nu)
        \nonumber \\
        && + {\Re}(H_4 \mathcal{M}_6^*) u_2^2 (u_1 u_2 - t_1 t_2 - s \hat s)
        - 4 {\Im}(H_4 \mathcal{M}_6^*) u_2^2 \epsilon(e^-,e^+,\anti \nu,\nu)
        \nonumber \\
K^{\prime}_8 &=& 4 {\Re}(F \mathcal{M}_8^*)(t_1 t_2 + u_1 u_2 - s \hat s)
                \epsilon(e^-,e^+,\anti \nu,\nu)
        \nonumber \\
        && + {\Im}(F \mathcal{M}_8^*) 
                [-(t_1 t_2 + u_1 u_2 - s \hat s)^2 + 2 t_1 t_2 u_1 (u_1 + u_2)]
        \nonumber \\
        && + 2 {\Re}(G \mathcal{M}_8^*) u_1 (t_1 t_2 + u_1 u_2 - s \hat s)
        + 8 {\Im}(G \mathcal{M}_8^*) u_1 \epsilon(e^-,e^+,\anti \nu,\nu)
        \nonumber \\
        && + {\Re}(H_1 \mathcal{M}_8^*) s u_1 (t_1 t_2 + u_1 u_2 - s \hat s)
        + 4 {\Im}(H_1 \mathcal{M}_8^*) s u_1 \epsilon(e^-,e^+,\anti \nu,\nu)
        \nonumber \\
        && + {\Re}(H_2 \mathcal{M}_8^*) \hat s u_1
                (t_1 t_2 + u_1 u_2 - s \hat s)
        + 4 {\Im}(H_2 \mathcal{M}_8^*) \hat s u_1 
                \epsilon(e^-,e^+,\anti \nu,\nu)
        \nonumber \\
        && + {\Re}(H_3 \mathcal{M}_8^*) u_1^2 (u_1 u_2 - t_1 t_2 - s \hat s)
        + 4 {\Im}(H_3 \mathcal{M}_8^*) u_1^2 \epsilon(e^-,e^+,\anti \nu,\nu)
        \nonumber \\
        && + {\Re}(H_4 \mathcal{M}_8^*) 
                [-(t_1 t_2 - s \hat s)^2 + u_1 u_2 (t_1 t_2 + s \hat s)]
        \nonumber \\
        && + 4 {\Im}(H_4 \mathcal{M}_8^*) (s \hat s - t_1 t_2)
                \epsilon(e^-,e^+,\anti \nu,\nu)
        \nonumber \\
K^{\prime}_{10} &=& 4 {\Re}(F \mathcal{M}_{10}^*)
                t_2 (u_1 + u_2) \epsilon(e^-,e^+,\anti \nu, \nu)
        \nonumber \\
        && - {\Im}(F \mathcal{M}_{10}^*)
                t_2 (s\hat s - t_1 t_2 - u_1 u_2) (u_1 - u_2)
        \nonumber \\
        && + 4 {\Re}(G \mathcal{M}_{10}^*) t_2 u_1 u_2
        + 2 {\Re}(H_1 \mathcal{M}_{10}^*) s t_2 u_1 u_2
        + 2 {\Re}(H_2 \mathcal{M}_{10}^*) \hat s t_2 u_1 u_2
        \nonumber \\
        && + {\Re}(H_3 \mathcal{M}_{10}^*) t_2 u_1 
                (s \hat s + u_1 u_2 - t_1 t_2)
        + 4 {\Im}(H_3 \mathcal{M}_{10}^*) t_2 u_1 
                \epsilon(e^-,e^+,\anti\nu,\nu)
        \nonumber \\
        && + {\Re}(H_4 \mathcal{M}_{10}^*) t_2 u_2
                (s \hat s + u_1 u_2 - t_1 t_2)
        - 4 {\Im}(H_4 \mathcal{M}_{10}^*) t_2 u_2 
                \epsilon(e^-,e^+,\anti\nu,\nu).
\end{eqnarray}

\end{document}